\documentclass[twocolumn]{aastex631}

\shorttitle{ALMA observations of GRS~1734-292}
\shortauthors{Michiyama et al.}
\graphicspath{{./}{figures/}}
\usepackage{amsmath}
\usepackage{comment}

\begin{document}

\title{ALMA Confirmation of Millimeter Time Variability in the Gamma-Ray Detected Seyfert Galaxy GRS~1734-292}

\correspondingauthor{Tomonari Michiyama}
\email{t.michiyama.astr@gmail.com}

\author[0000-0003-2475-7983]{Tomonari Michiyama}
\affiliation{Department of Earth and Space Science, Graduate School of Science, Osaka University, 1-1, Machikaneyama, Toyonaka, Osaka 560-0043, Japan}
\affiliation{National Astronomical Observatory of Japan, National Institutes of Natural Sciences, 2-21-1 Osawa, Mitaka, Tokyo, 181-8588}
\affiliation{Faculty of Information Science, Shunan University, 843-4-2 Gakuendai, Shunan,
Yamaguchi, 745-8566, Japan}

\author[0000-0002-7272-1136]{Yoshiyuki Inoue}
\affiliation{Department of Earth and Space Science, Graduate School of Science, Osaka University, 1-1, Machikaneyama, Toyonaka, Osaka 560-0043, Japan}
\affiliation{Interdisciplinary Theoretical \& Mathematical Science Program (iTHEMS), RIKEN, 2-1 Hirosawa, Saitama, 351-0198, Japan}
\affiliation{Kavli Institute for the Physics and Mathematics of the Universe (WPI), UTIAS, The University of Tokyo, 5-1-5 Kashiwanoha, Kashiwa, Chiba 277-8583, Japan}

\author{Akihiro Doi}
\affiliation{The Institute of Space and Astronautical Science, Japan Aerospace Exploration Agency, 3-1-1 Yoshinodai, Chuou-ku, Sagamihara, Kanagawa 252-5210,
Japan}
\affiliation{Department of Space and Astronautical Science, SOKENDAI, 3-1-1 Yoshinodai, Chuou-ku, Sagamihara, Kanagawa 252-5210, Japan}

\author{Tomoya Yamada}
\affiliation{Department of Earth and Space Science, Graduate School of Science, Osaka University, 1-1, Machikaneyama, Toyonaka, Osaka 560-0043, Japan}

\author{Yasushi Fukazawa}
\affiliation{Department of Physical Science, Hiroshima University, 1-3-1 Kagamiyama, Higashi-Hiroshima, Hiroshima 739-8526, Japan}
\affiliation{Hiroshima Astrophysical Science Center, Hiroshima University, 1-3-1 Kagamiyama, Higashi-Hiroshima, Hiroshima 739-8526, Japan}
\affiliation{Core Research for Energetic Universe (Core-U), Hiroshima University, 1-3-1 Kagamiyama, Higashi-Hiroshima, Hiroshima 739-8526, Japan}

\author{Hidetoshi Kubo}
\affiliation{Institute for Cosmic Ray Research, University of Tokyo, 5-1-5, Kashiwa-no-ha, Kashiwa, Chiba 277-8582, Japan}

\author{Samuel Barnier}
\affiliation{Department of Earth and Space Science, Graduate School of Science, Osaka University, 1-1, Machikaneyama, Toyonaka, Osaka 560-0043, Japan}
\affiliation{National Astronomical Observatory of Japan, National Institutes of Natural Sciences, 2-21-1 Osawa, Mitaka, Tokyo, 181-8588}








\begin{abstract}
GRS~1734-292 is a radio-quiet galaxy, exhibiting neither intense starburst nor jet activities. However, Fermi-LAT detected this object in the GeV band. 
The origin of non-thermal activity in this Seyfert galaxy is an intriguing question.
We report Atacama Large Millimeter/submillimeter Array (ALMA) observations of GRS~1734-292 at frequencies of 97.5, 145, and 225\,GHz.
These observations confirmed the millimeter excess within the central $\lessapprox100$\,pc region and its time variability based on two separate observations conducted four days apart.
The timescale of variability aligns with the light crossing time for a compact source smaller than  $<100$ Schwarzschild radius.
If we take into account the power-law synchrotron emission originating from the corona (i.e., the hot plasma located above the accretion disk), the  millimeter spectrum indicates the coronal magnetic field of $\approx10$\,G and the size of $\approx10$ Schwarzschild radius.
An alternative explanation for this millimeter emission could be synchrotron and free-free emission from disk winds (i.e., fast wide-opening angle outflows from the disk) with the size of $\approx10$\,pc, although it may be difficult to explain the fast variability.
Future millimeter observations with higher resolution ($<0\farcs01$) will enable the differentiation between these two scenarios. Such observations will provide insights into the acceleration sites of high-energy particles at the core of active galactic nuclei.
\end{abstract}

\keywords{
Astrophysical black holes (98),
Black hole physics (159), 
Black holes (162), 
Supermassive black holes (1663), 
Active galactic nuclei (16),
Seyfert galaxies (1447),
Particle astrophysics (96),
High energy astrophysics (739)
}


\section{Introduction}
The Fermi Gamma-ray Space Telescope has cataloged 6658 gamma-ray objects thus far \citep{Abdollahi2022ApJS..260...53A}. Although blazars make up the majority of these detections, Fermi has also unveiled new gamma-ray emitting populations, including starburst galaxies and radio galaxies \citep{Acciari2019ApJ...883..135A,Fukazawa2022ApJ...931..138F}. Intriguingly, recent observations \citep{Ackermann2012ApJ...747..104A,Wojaczy2015A&A...584A..20W} have shown that gamma rays have been detected not only in starburst galaxies and radio galaxies but also in unjetted active galactic nuclei (AGNs) known as radio-quiet Seyfert galaxies \citep{Seyfert1943ApJ....97...28S,Khachikian1974ApJ...192..581K,Hills1975Natur.254..295H}. Complementing this, a decade of data accumulation from the IceCube survey has identified a neutrino hotspot in the direction of the gamma-ray bright, radio-quiet Seyfert galaxy NGC~1068, with a significance level of 4.2 sigma \citep{IceCube2022Sci...378..538I}. The detection of gamma rays in Seyfert galaxies, coupled with a potential neutrino signal from NGC~1068, implies that radio-quiet Seyferts may constitute a substantial portion of the cosmic gamma-ray/neutrino sky. 
As the surface number density of Seyferts is approximately four orders of magnitude higher than that of blazars \citep{Luo2017ApJS..228....2L,Marcotulli2020ApJ...896....6M}, deciphering the origin of gamma-ray/neutrino emissions in radio-quiet Seyfert galaxies has emerged as a critical task in the field of high-energy astrophysics.

Although the detection of gamma rays in radio-quiet Seyfert galaxies represents a significant development in high-energy astrophysics, the source of these gamma rays remains enigmatic. One of the main challenges lies in the fact that various phenomena, including starbursts, jets, and central AGN activities, can all contribute to a galaxy's gamma-ray flux. This is particularly relevant in the case of well-known AGNs such as NGC~4945, NGC~1068, and Circinus, where it becomes difficult to disentangle the various sources of gamma radiation \citep{Lenain2010A&A...524A..72L,Acciari2019ApJ...883..135A,Michiyama2022ApJ...936L...1M}. Besides starbursts and jets, two other potential locations for particle acceleration also exist, which are the disk corona, the hot plasma situated above the disk, and the disk wind, fast wide-angle outflow emanating from the disk. 
In the context of this study, our primary emphasis centers on the investigation of the disk corona and the disk wind as potential locations for particle acceleration in Seyfer galaxies. 

The disk corona refers to a moderately optically-thick thermal plasma located above an accretion disk \citep{Galeev1979ApJ...229..318G}. It plays a role in producing X-ray photons in Seyfert galaxies by Comptonizing the photons emitted from the disk. The study of X-ray spectra allows us to derive crucial parameters associated with the coronal structure, including the coronal electron temperature ($T_{\rm e}$) and the Thomson scattering optical depth ($\tau$) \citep{Fabian2015MNRAS.451.4375F,Ricci2018MNRAS.480.1819R}.
In the case of nearby bright Seyfert galaxies, a typical spectral cut-off suggests a coronal temperature of $kT_{\rm e}\approx50~{\rm keV}$, where $k$ represents the Boltzmann constant. Additionally, the typical photon index of the X-ray emission
corresponds to $\tau\approx1$ \citep{Zdziarski10.1093,Ueda2014ApJ...786..104U}.
For a recent measurement of the  $kT_{\rm e}$ and $\tau$, refer to Figure 9 in \citet{Pal2023arXiv231018196P}.

It has been hypothesized that non-thermal electrons and protons present in these AGN coronae generate gamma-ray and neutrino emissions through inverse Compton (IC) scattering, hadronuclear ($pp$), and photomeson ($p\gamma$) interactions \citep[see e.g.,][]{Inoue2019ApJ...880...40I}.
As per the corona model proposed by \citet{Inoue2014PASJ...66L...8I}, emissions in the millimeter-band would exhibit characteristics of compact synchrotron self-absorption (SSA). The shape of the millimeter SSA is pivotal for estimating the corona's magnetic field (which is primarily related to SSA frequency) and its size (which is mainly related to SSA flux) \citep{Inoue2018ApJ...869..114I,Inoue2020ApJ...891L..33I,Michiyama2023PASJ...75..874M}.
These parameters play a crucial role in determining the flux of gamma rays and neutrinos. 
Notably, the Atacama Large Millimeter/submillimeter Array (ALMA) has already detected indications of non-thermal particles at the cores of several Seyfert galaxies \citep{Kawamuro2022ApJ...938...87K, Kawamuro2023ApJS..269...24K, Ricci2023ApJ...952L..28R}.

In this millimeter coronal synchrotron scenario, the inferred size of the corona is approximately $\approx10-100\,R_{\rm sh}$ ($R_{\rm sh}$ is Schwarzschild radius), which is significantly smaller than that suggested by other scenarios (e.g., wind model). 
Given its compactness and high-energy activity, various corresponding time variabilities are expected. 
A typically expected timescale is the light-crossing time, approximately  $\approx R_c/c$, which translates to about one-day variability for a black hole mass of $M_{\rm BH}=10^8\,M_{\odot}$.
Detecting such timescale variabilities would support the compact mm-wave emitting scenario, e.g., the corona, which is crucial for understanding the high-energy particle production processes in the coronae.

Another possible origin for the millimeter emission is a shock region within fast and massive disk winds. Detection of blue-shifted absorption features in UV and X-ray spectra have confirmed the existence of such winds \citep{Laha2021NatAs...5...13L}. It is proposed that these winds, upon interacting with the surrounding gas, could generate shocks strong enough to accelerate cosmic rays to relativistic energies. This acceleration process may, in turn, result in the production of gamma rays and neutrinos \citep{Wang2016NatPh..12.1116W, Ajello2021ApJ...921..144A}. The disk wind may contribute to the millimeter flux through synchrotron (and potentially free-free) emission from accelerated electrons \citep{Nims2015MNRAS.447.3612N}. 

To understand coronal and/or wind activities as an origin of high-energy particle, it is crucial to analyze millimeter emissions in Seyfert galaxies that have been detected in gamma rays, particularly those lacking prominent starbursts and jets. This is imperative to understand the origins of particle acceleration in the nuclei of Seyfert galaxies.
Among the gamma-ray sources listed in the latest Fermi catalog (Fermi LAT 12-Year Point Source Catalog), we have identified GRS~1734-292 (AX~J1737.4-2907) as an optimal target due to its significant gamma-ray flux, coupled with an absence of intense starburst and jet activity. This paper presents ALMA observations of GRS~1734-292 and investigates millimeter emissions originally either from the corona or disk winds.
This paper is structured as follows: Section~\ref{sec:GRS} explains the target GRS~1734-292 and Section~\ref{sec:ALMA} shows the ALMA observations. We argue possible millimeter origin in Section~\ref{sec:discussion}.
This paper assumes a cosmology with $H_0 = 70$~km~s$^{-1}$~Mpc$^{-1}$ and $\Omega_{\rm m}=0.3$.

\section{GRS~1734-292}\label{sec:GRS}
\subsection{Fermi detection of GRS~1734-292}
In order to identify Seyfert galaxies detected in gamma rays, two catalogs were utilized: the Swift BAT 105-Month Hard X-ray Survey catalog \citep{Oh2018ApJS..235....4O,Baumgartner2013ApJS..207...19B} and the Fermi LAT 12-Year Point Source Catalog (4FGL-DR3)\footnote{\url{https://fermi.gsfc.nasa.gov/ssc/data/access/lat/12yr_catalog/}} \citep{Abdollahi2022ApJS..260...53A}. 
The source named as 4FGL~J1737.1-2901 is the gamma-ray candidate of GRS~1734-292, which has the integral photon flux of $(9.8038\pm1.7733)\times10^{-10}\,\rm{ph\,cm^{-2}\,s^{-1}}$ (1 to 100\,GeV) and the energy flux of $(6.3935\pm2.5598)\times10^{-12}\,\rm{erg\,cm^{-2}\,s^{-1}}$ (100\,MeV to 100\,GeV).
The detection significance over the 100~MeV to 1~TeV band is 5.6$\sigma$.
GRS 1734-292 at $z=0.0214$ is the fifth nearby galaxy among Fermi-detected Seyferts in the order of 
Circinus Galaxy, 
NGC~4945, 
NGC~4151, and 
NGC~1068  
(e.g., \citealt{Lenain2010A&A...524A..72L, Hayashida2013ApJ...779..131H, Wojaczy2017ApJ...849...97W, Peretti2023arXiv230303298P}).

There are multiple reasons why the origin of the gamma-ray emissions from GRS~1734-292 cannot be attributed to starbursts or jet activity. Firstly, GRS~1734-292 exhibits a significantly low infrared luminosity of less than $<10^{10.41}\,L_{\odot}$ \citep{Shimizu2016MNRAS.456.3335S}. It does not adhere to the established correlation between infrared and gamma-ray luminosity observed in starburst galaxies, thus ruling out starburst activity as the primary source \citep{Ajello2020ApJ...894...88A}.
GRS~1734-292 also does not align the known correlation between the centimeter and gamma-ray luminosity of radio galaxies \citep{Inoue2011ApJ...733...66I, Fukazawa2022ApJ...931..138F}. Indeed, GRS~1734-292 exhibits a very bright gamma-ray luminosity compared to its 1.4\,GHz  radio flux of 63\,mJy
, which corresponds to a radio luminosity of $L_{\rm 1.4GHz}=7\times10^{22}\,{\rm W Hz^{-1}}$ \cite{Marti1998A&A...330...72M}.
Even considering the possibility of strong relativistic beaming effects, the gamma-ray luminosity of GRS~1734-292 remains challenging to explain. Consequently, GRS~1734-292 represents an important target for revealing the mechanisms of gamma-ray production in a ``pure" Seyfert galaxy.

\subsection{Details of GRS~1734-292}
GRS~1734-292 was first identified as a source of hard X-ray emissions emanating from the direction of the Galactic Center \citep{Sunyaev1990IAUC.5123....2S}. 
Follow-up optical spectroscopic observations unveiled strong and broad emission lines, attesting to the presence of a type-I Seyfert active galactic nucleus (AGN) nestled within GRS~1734-292 at a redshift of $z=0.0214$ \citep{Marti1998A&A...330...72M}.
The black hole mass is measured as
$\log{(M_{\rm BH}\,[M_{\odot}])}=8.50\pm0.50$ (see the Appendix in \citealt{Tortosa2017MNRAS.466.4193T} for details), corresponding Eddington luminosity as $\log{(L_{\rm edd} \,[\rm erg\,s^{-1}])}\approx46.60$.
The bolometric luminosity, corrected according to \citet{Marconi2003ApJ...589L..21M}, is estimated as $\log{(L_{\rm bol}^{\rm corr} \,[\rm erg\,s^{-1}])}\approx45.14$ \citep{Molina2019MNRAS.484.2735M}.
This results in an Eddington ratio of $\log{(L_{\rm bol}^{\rm corr}/L_{\rm edd})}=-1.46\pm0.50$.
The centimeter radio counterparts of GRS~1734-292 were identified as a pair of weak synchrotron jets, with each jet extending approximately $5\farcs$ \citep{Marti1998A&A...330...72M}. Accoring to \citet{Tortosa2017MNRAS.466.4193T}, 
the 0.1-100 GHz radio luminosity is $L_{\rm 0.1-100 GHz}\approx7\times10^{39}\,{\rm erg\,s^{-1}}$, 
and the 0.5-4.5 keV X-ray luminosity is $L_{\rm 0.5-4.5\,keV}=1\times10^{44}\,{\rm erg\,s^{-1}}$ \citep{Marti1998A&A...330...72M}, resulting in a radio loudness of $L_{\rm 0.1-100 GHz}/L_{\rm 0.5-4.5\,keV}=7\times10^{-5}$, classifying it as a radio-quiet AGN \citep{Laor2008MNRAS.390..847L}.
The INTEGRAL satellite precisely measured the hard X-ray spectrum of GRS~1734-292 \citep{DiCocco2004A&A...425...89D, Sazonov2004A&A...421L..21S}.
For instance, \citet{Tortosa2017MNRAS.466.4193T} reported a temperature of the coronal plasma of $kT_{\rm e}=11.9^{+1.2}_{-0.9}$\,keV and an optical depth of $\tau=2.98^{+0.16}_{-0.19}$ assuming a slab geometry.
\citep{Pal2023arXiv231018196P} have also measured the $kT_{\rm e}$ and $\tau$ of GRS~1734-292.
We have utilized the value measured in \citet{Tortosa2017MNRAS.466.4193T} since it corresponds to a specific case study of GRS~1734-292.
According to \citet{Pal2023arXiv231018196P}, the plasma temperature of $kT_{\rm e}=11.9^{+1.2}_{-0.9}$\,keV in \citet{Tortosa2017MNRAS.466.4193T} and $kT_{\rm e}=20^{+4}_{-2}$\,keV in \citet{Pal2023arXiv231018196P}, are slightly lower than the mean value among similar AGN populations.

\section{ALMA observations} \label{sec:ALMA}
\begin{figure*}[ht!]
\epsscale{1.1}
\plotone{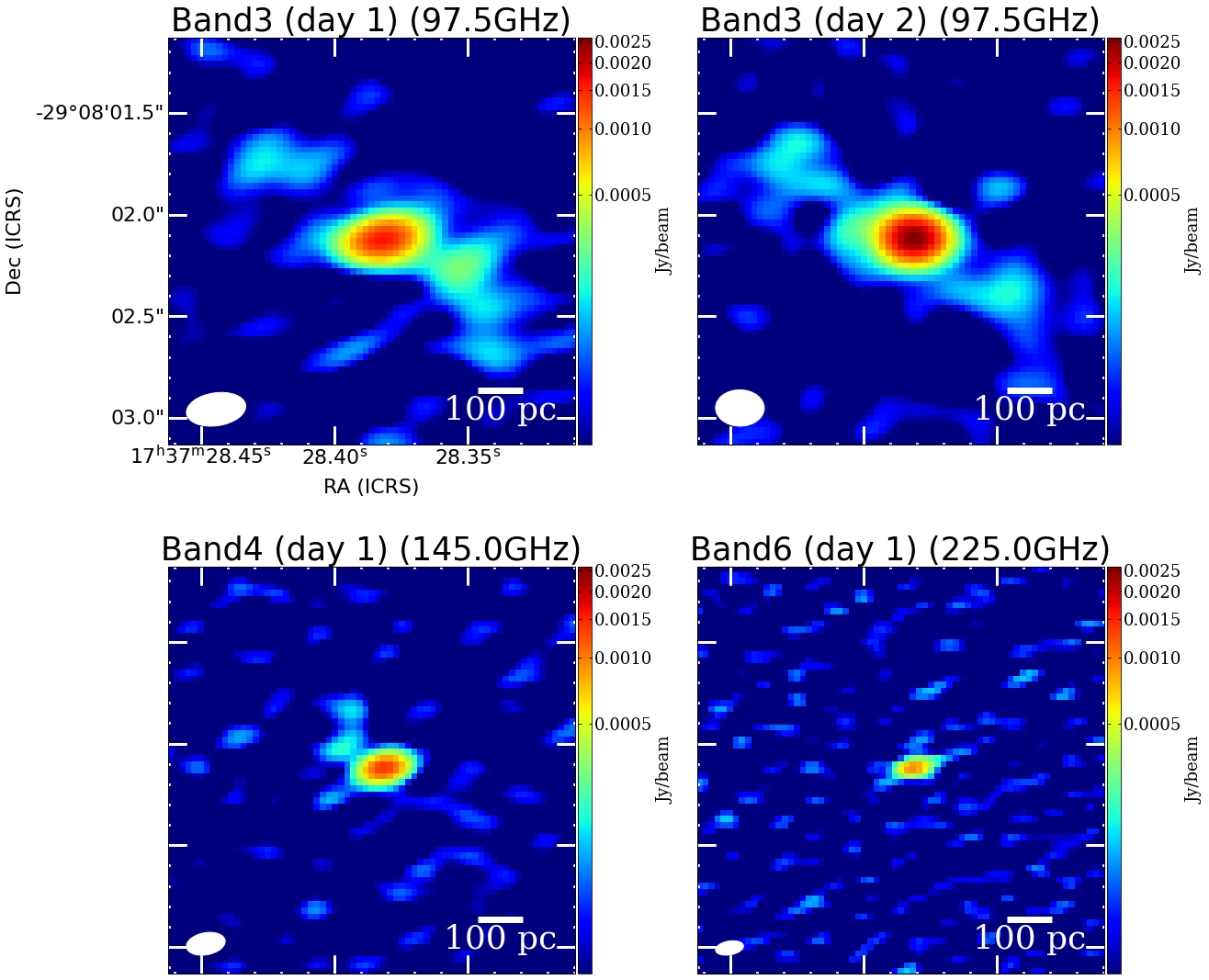}
\caption{The ALMA map of GRS 1734-292 at 97.5\,GHz, 145\,GHz, and 225\,GHz. The synthesized beam is represented by the white ellipse in the lower left corner, and the white bar in the lower right corner represents the scale bar of 100 pc. \label{fig:ALMA}}
\end{figure*}

\begin{figure}[ht!]
\epsscale{1.0}
\plotone{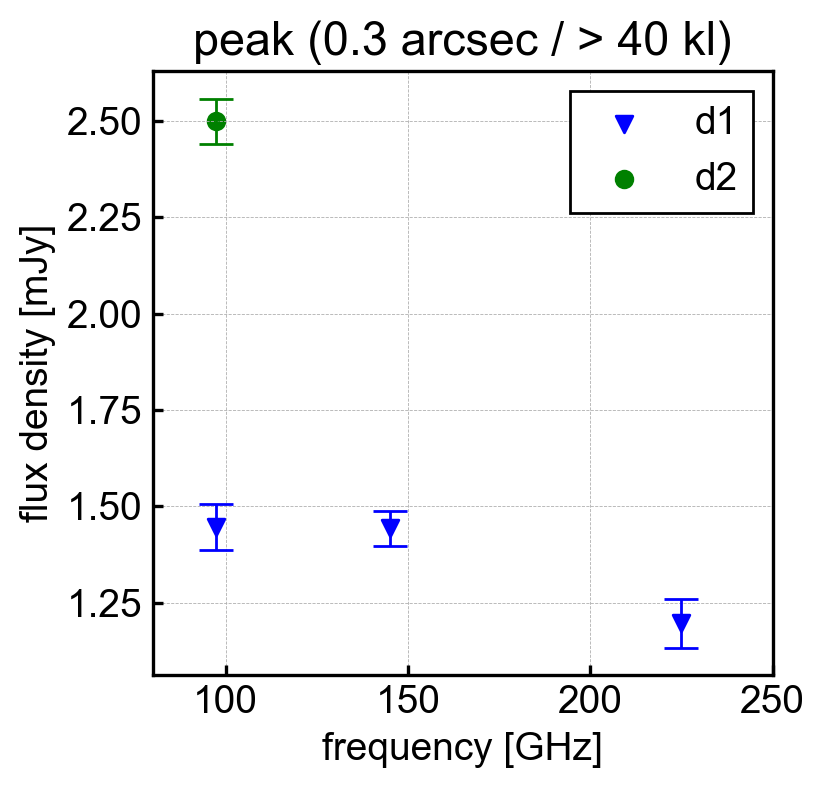}
\caption{The peak flux density of GRS 1734-292 based on each beam matched ($0\farcs3$) and uvclipped ($>40\,{\mathrm k\lambda}$) ALMA image.  \label{fig:ALMA_SED}}
\end{figure}

During the ALMA Cycle 6 survey (ALMA project code 2018.1.00576.S), we observed GRS~1734-292 on August 14th and 18th, 2019, using Band 3, 4, and 6 receivers (Table~\ref{tab:ALMA}). 
Using this data set as one target among statistical analysis (the BAT index is 896), \citet{Kawamuro2022ApJ...938...87K} confirmed that this target is on the typical relation between millimeter and X-ray luminosity.
In each execution block, the amplitude and phase calibrators are J1924-2914 and J1744-3116 and the quasar J1752-2956 was observed as a check source.
The raw visibility data was calibrated using {\tt CASA} \citep{McMullin_2007ASPC..376..127M,CASA2022arXiv221002276T} by running the provided calibration scripts.
The continuum images were generated using {\tt clean} in {\tt CASA} with a cell size of 0\farcs03, an image size of 3000 pixels, and Briggs weighting with a robust parameter of $-0.2$. 
Clean masks were determined by the automatic masking loop (sidelobethreshold = 2.0, noisethreshold = 3.0, lownoisethreshold = 2.0, minbeamfrac = 0.3, growiterations = 75, and negativethreshold = 0.0).
Figure~\ref{fig:ALMA} shows millimeter images at 97.5, 145, and 225\,GHz.
Table~\ref{tab:ALMA_flux} represents the image qualities such as the size of the synthesized beam, the noise level (rms), and the peak flux density ($F_{\rm peak}$).
For the purpose of comparing the millimeter flux at Band~3, 4, and 6, we also made beam matched ($0\farcs3$) and uv-clipped ($>40\,{\mathrm k\lambda}$) images (hearafter matched images) (Figure~\ref{fig:ALMA_SED}).
The $F_{0\farcs3}^{\rm peak}$ in Table~\ref{tab:ALMA_flux} represents the peak flux in the matched images and the $F_{0\farcs3}^{\rm imfit}$ is the total flux measured by the 2D Gaussian fit ({\tt imfit} task in {\tt CASA}).
The 97.5\,GHz flux density is $F_{0\farcs3}^{\rm peak}=1.45\pm0.06$\,mJy and $=2.5\pm0.06$\,mJy on August 14 and 18, 2019 respectively, corresponding to a 70\% variability in four days. 

\begin{deluxetable*}{lccc}
\tablecaption{ALMA execution blocks summary for GRS~1734-292 \label{tab:ALMA}}
\tablehead{
\colhead{Date} & \colhead{Band} & \colhead{Freq.} & \colhead{Baseline (MRS)} \\
\colhead{start, end (UTC)} & \colhead{} & \colhead{GHz} & \colhead{m, m (\farcs)} }
\decimalcolnumbers
\startdata
2019-08-14 22:06, 2019-08-14 22:27 & B6 & 224, 226, 240, 242 & 41.4, 3637.7 (1\farcs7) \\
2019-08-14 22:29, 2019-08-14 22:45 & B3 (d1) & 90.521, 92.416, 102.521, 104.479 & 41.4, 3737.3 (4\farcs3) \\
2019-08-14 22:46, 2019-08-14 23:05 & B4 & 138.021, 139.916, 150.021, 151.979 & 41.4, 3637.6 (2\farcs9) \\
2019-08-18 22:50, 2019-08-18 23:08 & B3 (d2) & 90.521, 92.416, 102.521, 104.479 & 41.4, 3637.3 (4\farcs1) \\
\enddata
\end{deluxetable*}

\begin{deluxetable*}{lcccccc}
\tablecaption{Summary of the ALMA image qualities and the flux measurements \label{tab:ALMA_flux}}
\tablehead{
\colhead{Band} & \colhead{Freq.} & \colhead{Beam} & \colhead{rms}  & \colhead{$F_{\rm peak}$} & \colhead{$F_{0\farcs3}^{\rm total}$} & \colhead{$F_{0\farcs3}^{\rm peak}$} \\
\colhead{} & \colhead{GHz} & \colhead{$\farcs\times\farcs$} & \colhead{mJy\,beam$^{-1}$}  & \colhead{mJy\,beam$^{-1}$} & \colhead{mJy} & \colhead{mJy}
}
\decimalcolnumbers
\startdata
B3 (d1) & 97.5 & $0.29\times0.16$ & 0.038 & 1.6 & $2.2\pm0.1$ & $1.45\pm0.06$\\
B3 (d2) & 97.5 & $0.24\times0.18$ & 0.04 & 2.5 & $3.1\pm0.1$ & $2.5\pm0.06$\\
B4 & 145 & $0.19\times0.11$ & 0.036 & 1.4 & $1.9\pm0.1$ & $1.44\pm0.04$\\
B6 & 225 & $0.14\times0.07$ & 0.046 & 0.9 & $1.6\pm0.1$ & $1.2\pm0.06$
\enddata
\end{deluxetable*}

\section{Discussion} \label{sec:discussion}

\subsection{Time variability}\label{sec:time}
\begin{figure}[ht!]
\epsscale{1.0}
\plotone{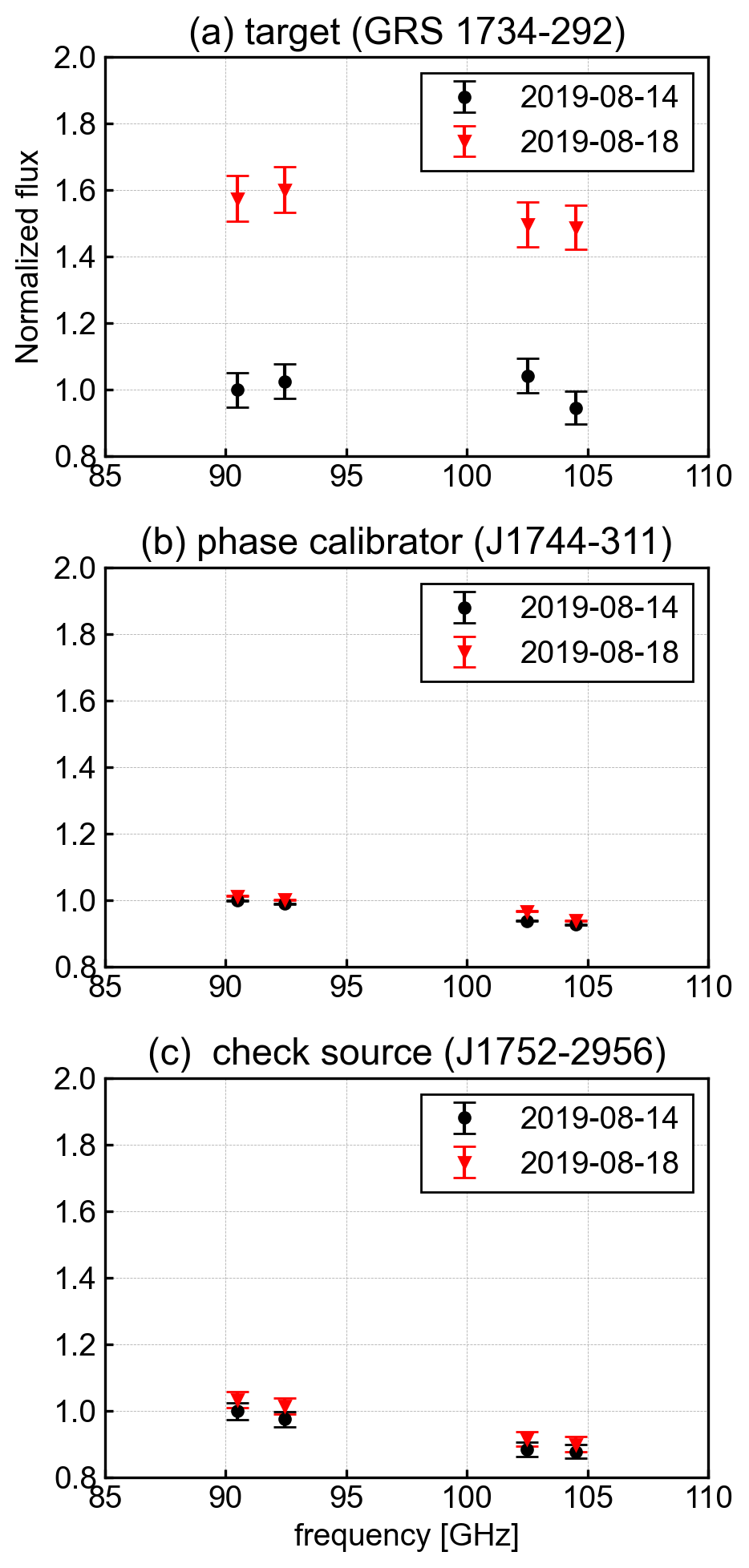}
\caption{The relative target flux for each spw in multi-epoch Band~3 observations for (a) target, (b) phase calibrator, and (c) check source, respectively. \label{fig:cal}}
\end{figure}

In Section~\ref{sec:ALMA}, we present the ALMA observational results, indicating $\approx70$~\% variability at 97.5\,GHz over four days. Considering that the amplitude calibration error of the ALMA design can be as high as 5~\% at Band~3\footnote{\url{https://almascience.nao.ac.jp/documents-and-tools/cycle6/alma-technical-handbook}}, this time variability is considered reliable. Furthermore, we have confirmed the presence of time variability in each spectral window (spw), ruling out spurious effects in specific spectral windows (see Figure~\ref{fig:cal}a). We have also investigated the flux density of the phase calibrator (J1744-3116) and the check source (J1752-2956) and found no systematic flux variability (see Figure~\ref{fig:cal}b and c). Finally, we have verified that the extended component seen in the Band 3 images (see Figure~\ref{fig:ALMA}) maintains consistent flux between the two observation epochs. Therefore, we conclude that the time variability observed in the nucleus of this target source is significant.

The size of the emitting region is smaller than the light-crossing timescale expressed as
\begin{equation}
R<c\Delta t,
\end{equation}
where $\Delta t\approx4$\,days is the time variablitiy timescale and $c$ is the speed of light. 
Figure~\ref{fig:cal} shows the confirmation of day-scale time variability at $\approx100$\,GHz.
By assuming that this timescale corresponds to the light-crossing timescale around the central black hole with a mass of $M_{\rm BH}\approx3\times10^8\,M_\odot$ \citep{Tortosa2017MNRAS.466.4193T}, we can infer a potential size for the emitting region as $R\lesssim100R_{\rm s}$, where $R_{\rm s}$ is the Schwarzschild radius. This size argument based on time variability suggests a compact source region, e.g., the corona. 
Considering that the alternative potential sources, such as extended (e.g., $>100\,R_{\rm s}$) jet synchrotron and dust blackbody, can not exhibit daily time variability, it is plausible to conclude that the coronal synchrotron emission makes a significant contribution to the millimeter fluxes detected by ALMA.

\subsection{Millimeter excess} \label{sec:mm}
\begin{figure}[ht!]
\epsscale{1.0}
\plotone{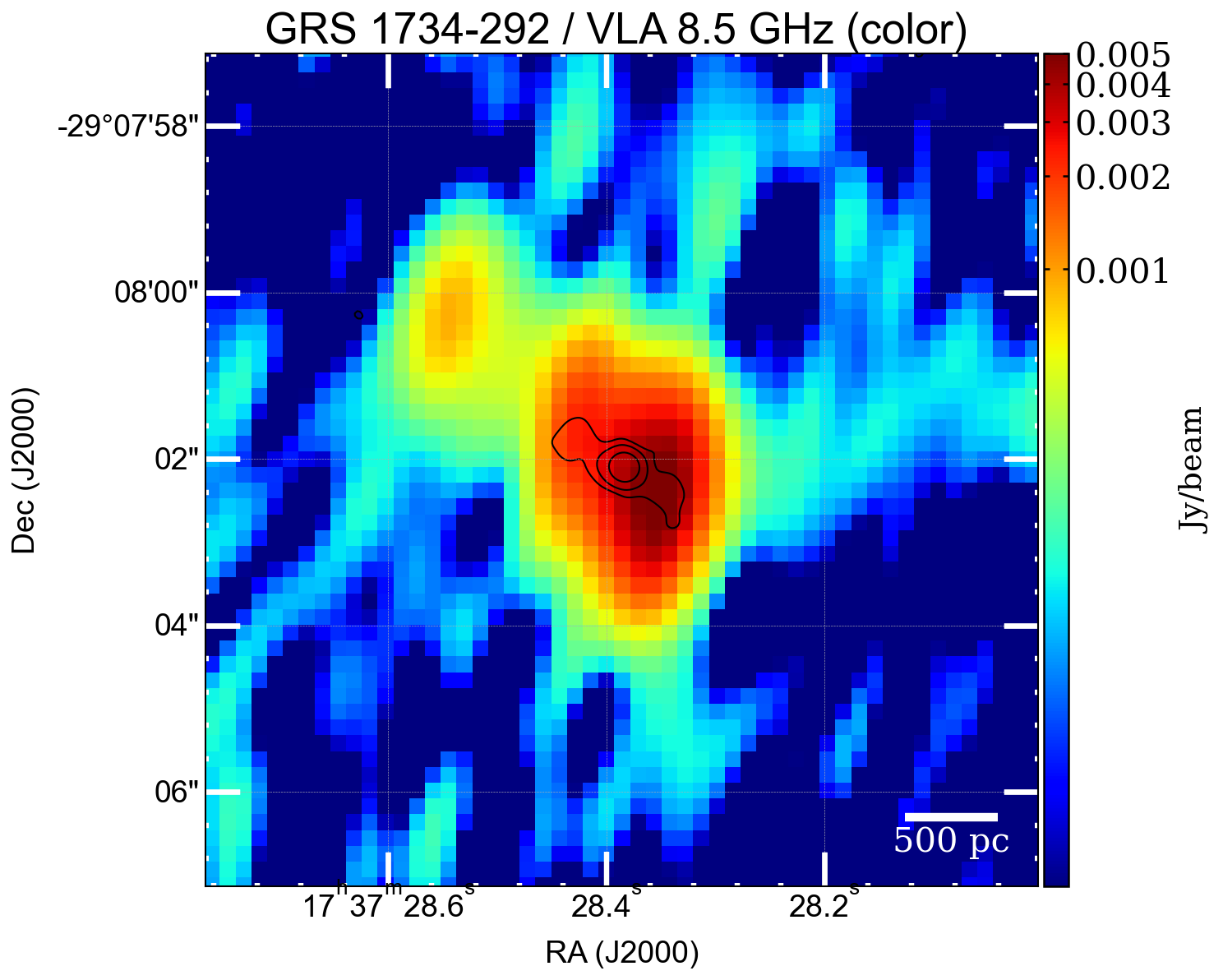}
\caption{The 8.5\,GHz continuum map obtained by VLA (color) and ALMA 97.5\,GHz continuum: (3, 9, and 27)$\times0.04$\,mJy\,beam$^{-1}$ contour. \label{fig:ALMA_VLA}}
\end{figure}

\begin{figure}[ht!]
\epsscale{1.0}
\plotone{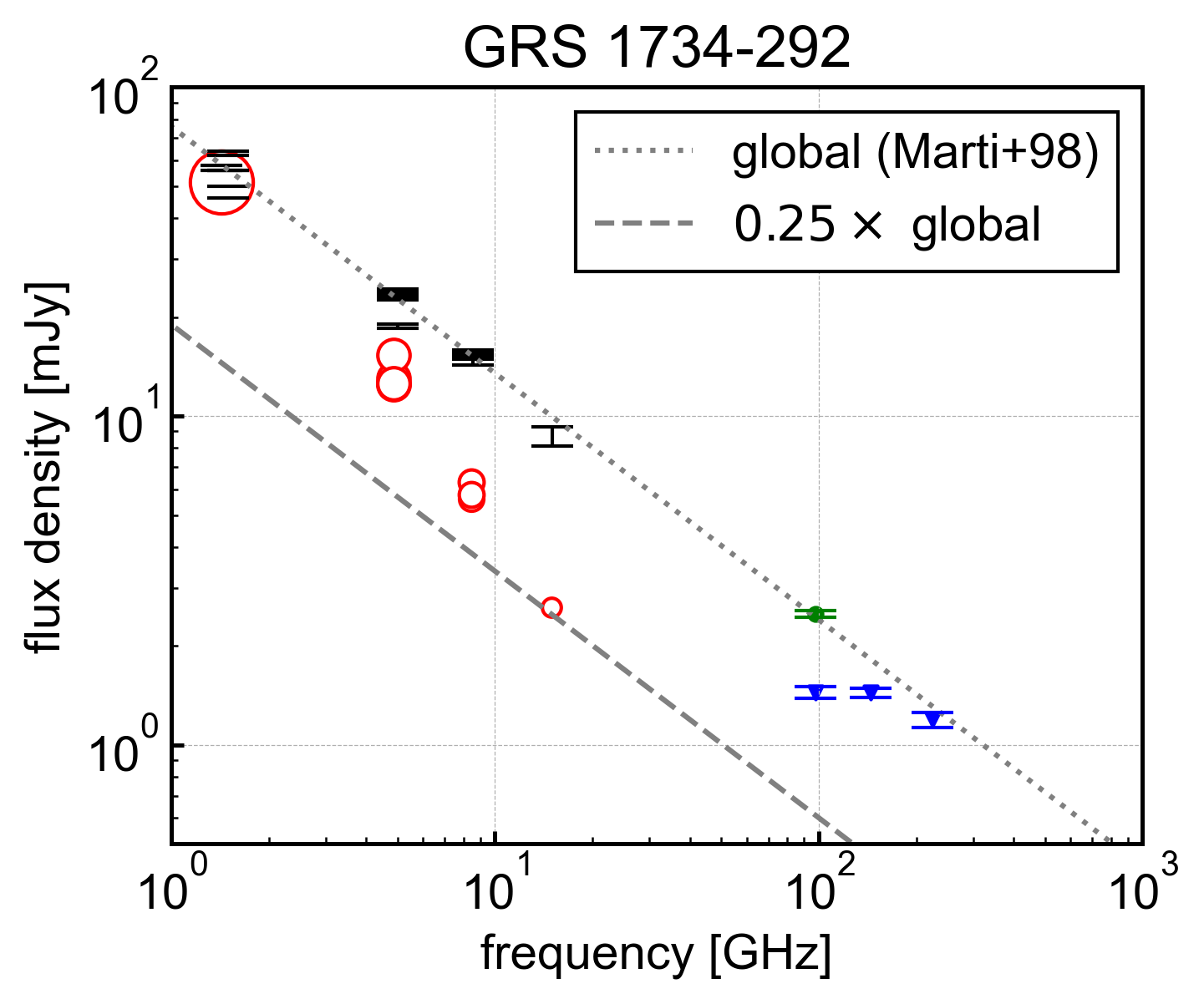}
\caption{The black bars represent the flux of global jet components following the gray line $S_{\nu}/{\rm mJy}=76\nu^{-0.75}$ \citep{Marti1998A&A...330...72M}. The red circles show the peak flux density at the ALMA 97.5\,GHz continuum peak position. The size of the circles corresponds to the synthesized beam of each image. The typical beamsize is 6\farcs5 at 1.4\,GHz, 1\farcs7 at 4.8\,GHz, 1\farcs0 at 8.5\,GHz, and 0\farcs6 at 15\,GHz, respectively. The green and blue symbols are the same as in Figure~\ref{fig:ALMA_SED}. \label{fig:radio_SED}}
\end{figure}

Considering the detection of a double-sided jet in the centimeter bands \citep{Marti1998A&A...330...72M}, it is reasonable to assume that the jet synchrotron emission would also contribute to the millimeter flux, potentially contaminating it.
Figure~\ref{fig:ALMA_VLA} shows the Very Large Array (VLA) 8.5\,GHz map\footnote{The calibrated VLA continuum image fits files were downloaded in NRAO VLA Archive Survey (\url{https://www.vla.nrao.edu/astro/nvas/}).}.
The alignment between the extended jet-like structure observed in the ALMA map and the more extended jet seen in the VLA\,8.5 GHz map suggests that spatially distinguishing jet contamination may be possible.
The peak position between ALMA millimeter and VLA 8.5\,GHz differs by $\approx1\farcs$. 
This suggests that spatially extended jet components dominate at 8.5\,GHz, while the nuclear region has a millimeter excess that does not originate from the extended jet. It is important to note that the spatially integrated flux density at VLA 8.5\,GHz is  $\approx15$\,mJy, while the emission within the ALMA beam is expected to be less than 6\,mJy. Similar beamsize effects are seen at other wavelengths as well.
Figure~\ref{fig:radio_SED} shows that the global jet spectrum follows $S_{\mu}\approx76(\mu/{\rm GHz})^{-0.75}$ (gray dotted line).
We investigate spatially unresolved components from the archival continuum image (red circles) and find that the ALMA beamsize (0\farcs3) can cover less than 25\% of the global jet component (gray dashed line).
For instance, the total flux at 14\,GHz is $\approx8.7$\,mJy but the emission around the ALMA peak position is $\lesssim2.6$\,mJy.
Because the ALMA flux exceeds the gray dashed line in Figure~\ref{fig:radio_SED}, the main millimeter source is unlikely to be the radio jet.
Furthermore, the nearly flat spectral index between Band 3 and 7 contradicts the expected spectral index of 3--4 for a dust gray body. Therefore, we can confirm that the ALMA observations reveal a millimeter excess.

\subsection{Corona}
\begin{figure}[ht!]
\epsscale{1.0}
\plotone{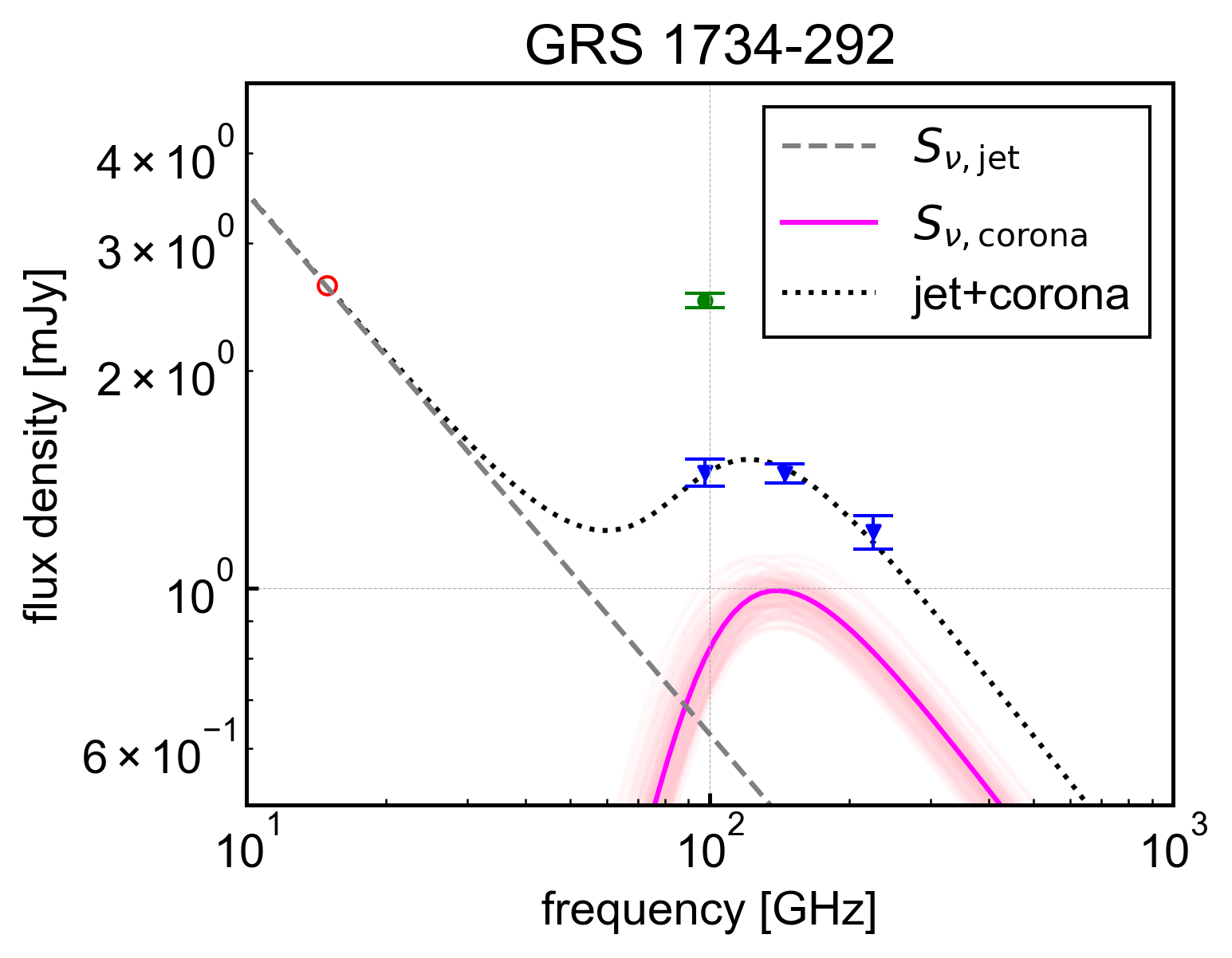}
\caption{The magenta line shows the coronal SSA ($S_\mathrm{\nu_{\rm SSA}}\approx1.5$\,mJy and $\nu_{\rm SSA}\approx110$\,GHz in equation~\ref{eq:corona}) and the gray dashed line shows the jet contamination ($f_{\rm jet}\approx0.26$ in equation~\ref{eq:jet}). The black dotted line shows the sum of jet and coronal emissions. \label{fig:Bay}}
\end{figure}

It is widely believed that X-ray in Seyfert galaxies is primarily due to the Comptonized accretion disk photons from moderately thick thermal plasma, i.e., coronae, above an accretion disk \citep{Galeev1979ApJ...229..318G}.
We can model the coronal radio emission using synchrotron self-absorption (SSA) as
\begin{equation}\label{eq:corona}
S_\mathrm{\nu,\rm{corona}}=S_\mathrm{\nu_{SSA}}\left( \frac{\nu}{\nu_\mathrm{SSA}} \right)^{5/2}\left\{1-\exp\left[\left(-\frac{\nu}{\nu_\mathrm{SSA}} \right)^{-(p+4)/2}\right]\right\},
\end{equation}
where $S_\mathrm{SSA}$ is the normalizations, $\nu_\mathrm{SSA}$ is the SSA frequency, and $p$ is the slope of the electron-energy spectrum (the slope of $\alpha_{\rm corona}=(1-p)/2$ in the SED).
Determining $S_\mathrm{SSA}$ and $\nu_\mathrm{SSA}$ provides us with the means to estimate crucial parameters such as the coronal magnetic field strength ($B$) and size ($R$). For instance, one can establish relationships like $B\propto \nu_\mathrm{SSA} S_\mathrm{SSA}^{-0.1}$ and $R\propto \nu_\mathrm{SSA}^{-1} S_\mathrm{SSA}^{0.5}$ when $\delta=2.7$. The detailed mathematical expressions for these relationships can be found in the work by \citet{Chaty2011A&A...529A...3C}.
Using the Bayesian parameter estimation technique developed by \citet{Foreman-Mackey2013PASP..125..306F}\footnote{\url{https://emcee.readthedocs.io/en/v2.2.1/user/line/}},
we estimate $S_\mathrm{\nu_{\rm SSA}}$ and $\nu_{SSA}$.
We also add contamination from the global jet synchrotron (see Figure~\ref{fig:radio_SED} and \citealt{Marti1998A&A...330...72M}) by considering the jet fraction ($f_{\rm jet}$) as the free-parameter during the Bayesian analysis by 
\begin{equation}\label{eq:jet}
S_\mathrm{\nu,\rm{jet}}=f_{\rm jet}\times\left(\frac{76}{\rm mJy}\right)\left(\frac{\nu}{\rm GHz}\right)^{-0.75}. 
\end{equation}
We use ALMA Band~3, 4, and 6 flux in day-1 and VLA 15\,GHz un-resolved flux as the input data, and the fitting results are shown in Figure~\ref{fig:Bay}.
The estimated parameters are ($S_\mathrm{\nu_{\rm SSA}}$, $\nu_{\rm SSA}$, $f_{\rm jet}$) = ($1.5^{+0.08}_{-0.08}$\,mJy, $110^{+13}_{-6}$\,GHz, $0.26^{+0.02}_{-0.02}$) assuming $p=2.7$, where the errors are calculated based on the 16th, 50th, and 84th percentiles of the posterior distributions.

Following \citet{Inoue2020ApJ...891L..33I}, the derived $S_\mathrm{\nu_{\rm SSA}}\approx1.5$\,mJy and $\nu_{\rm SSA}\approx110$\,GHz correspond to the coronal magnetic field $B\approx10$\,G and the size $R\approx10\,R_{\rm s}$, where we assume a temperature of the coronal plasma of $kT_{\rm e}=11.9^{+1.2}_{-0.9}$\,keV and an optical depth of $\tau=2.98^{+0.16}_{-0.19}$  (suggested by hard X-ray modeling; \citealt{Tortosa2017MNRAS.466.4193T}).
The detailed calculation processes are given in \citet{Inoue2014PASJ...66L...8I}.
Some numerical simulations predict stronger magnetic fields in the coronal region than our result \citep{Liska2022ApJ...935L...1L}. These simulations often presuppose strong large-scale poloidal magnetic fields to reproduce powerful jets seen in radio-loud AGNs successfully. However, GRS~1734-292 is a radio-quiet AGN, which lacks powerful jet activity. In this case, a toroidal magnetic field configuration is preferred, resulting in weaker coronal magnetic fields than those associated with large-scale poloidal magnetic fields due to the Parker instability \citep{Parker1955ApJ...121..491P, Parker1966ApJ...145..811P,Takasao2018ApJ...857....4T,Liska2022ApJ...935L...1L}, which is consistent with our estimate \citep[see][for detailed discussions]{Inoue2024arXiv240107580I}.

\subsection{Disk wind}
\begin{figure}[ht!]
\epsscale{1.0}
\plotone{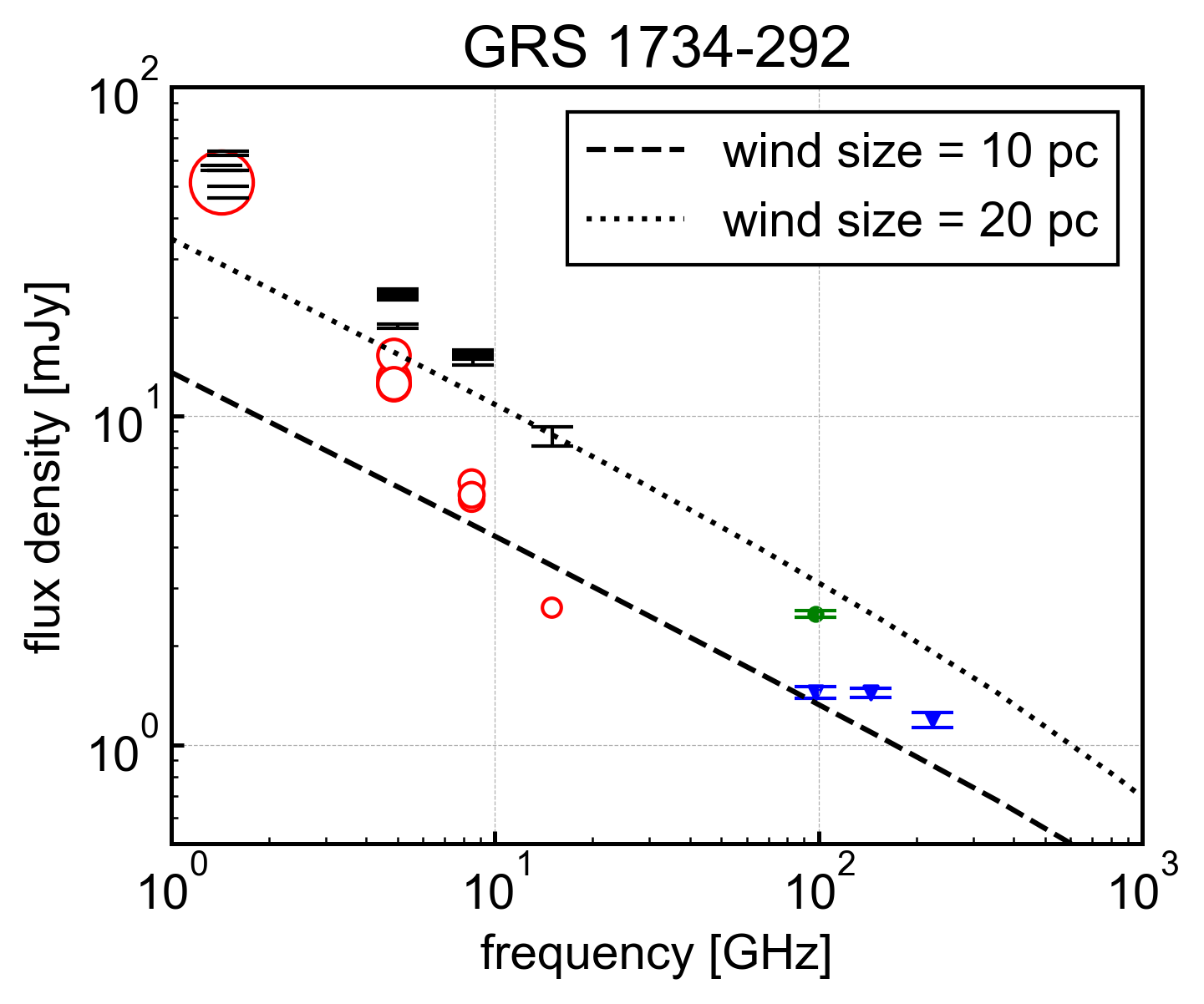}
\caption{The lines show the predicted flux using the wind model \citep{Nims2015MNRAS.447.3612N} with the size of 10\,pc (black dashed line) and 20\,pc (gray dotted line). The other symbols are the same as in Figure~\ref{fig:ALMA_VLA}. \label{fig:wind}}
\end{figure}

Fast and powerful winds are likely to be ejected from accretion disks, which are observed as blue-shifted absorption features in UV and X-rays \citep{Laha2021NatAs...5...13L}.
In GRS~1734-292, \citet{Tortosa2017MNRAS.466.4193T} identified disk winds with a velocity of 9500\,km\,s$^{-1}$ ($\approx0.03c$) based on observations of Fe~XXVI ions (rest-frame energy: 6.966\,keV).
In addition to the contribution from coronal emission, it is also plausible that the radio emission originating from the wind could contribute to the observed millimeter excess.
The radio flux, in terms of synchrotron and free-free emission from accelerated electrons, is dependent on the volume of the emitting region, which corresponds to the size of the wind.
If the size of the wind is $\approx10$\,pc, the predicted emission at $\approx100$\,GHz is $\approx2$\,mJy (Figure~\ref{fig:wind}) for winds with the velocity of $0.03c$ and the AGN luminosity of $1.45\times10^{45}$\,erg\,s$^{-1}$ after 2--10\,keV bolometric correction \citep{Marconi2004MNRAS.351..169M}.
We assume a power-law electron distribution, 1\% of the shock energy being transformed into relativistic electrons, and a magnetic field typical of the interstellar medium (See \citealt{Nims2015MNRAS.447.3612N} for the detailed formulation).
This means that compact wind ($<1$\,pc) cannot explain the ALMA continuum flux measurements from archival 0\farcs3 images.
Or otherwise, a compact wind ($<1$\,pc) could be responsible for the emission if the magnetic field were $>10$\,mG.
But a magnetic field that amplifies so much is unusual.
Based on the inability of the wind model to account for millimeter time variability, it is unlikely that wind emission is the primary source of the observed millimeter excess. Consequently, the disk wind can be considered a potential ``contamination" of the millimeter coronal synchrotron self-absorption (SSA). If the wind model is true, future ALMA observations with even higher resolution  ($<0\farcs05$) could resolve the structure and provide more insights.
Further details regarding the calculation of the radio spectral energy distribution regarding the disk wind will be published in a separate paper (Yamada et al.under review at ApJ).

\subsection{Jet base}
A very compact region at the jet base can also become a possible origin of the millimeter flux. The jet power in GRS~1734-292 has been calculated to be
$P_{\rm jet}=1.5\times10^{43}\,{\rm erg\,s^{-1}}$, based on a scaling relation between jet power and radio power \citep{Birzan2008ApJ...686..859B}.
At the jet base, the jet is expected to be dominated by the Poynting flux ($u_B \gg u_e$), thus we expect $P_{\rm jet} \approx P_B = 4\pi R_{\rm c}^{2} c u_{\rm B}$, where $u_e$ and $u_B$ are the energy densities of electrons and magnetic fields, respectively, and $P_B$ is the magnetic power.
Assuming $R_{\rm c} = 100\,R_{\rm s}$, we have the magnetic field strength of $B_{\rm jet} \approx 3\,{\rm G}$ at the jet base. 
Given the observed frequency at $\nu_{\rm syn} = 100\,{\rm GHz}$ and $B_{\rm jet} \approx 3\,{\rm G}$, the corresponding electron Lorentz factor is $\gamma \approx 80$.
The required electron number at this Lorentz factor becomes $N_{\rm e}(\gamma = 80) \approx 3\times10^{49}$ electrons, considering synchrotron emissivity. With an electron spectral index of $-2.5$ as indicated in the radio spectrum, the electron energy density is $u_e \approx 1\,{\rm erg\,cm^{-3}}$, while $u_B \approx 0.5\,{\rm erg\,cm^{-3}}$.
This result $(u_B\sim u_e)$ contradicts with the initial assumption of $u_B\gg u_e$.  Therefore, the jet-base scenario might be untenable.
Observationally, simultaneous X-ray and radio observations are the key to proving that X-rays and radio come from the same source.

\section{Summary}
GRS~1734-292, a galaxy known for its detection of gamma-ray emission, is an ideal candidate for investigating the regions where gamma-rays are produced in radio-quiet AGN because it is not contaminated either by intense starbursts and jets. 
ALMA confirmation of the millimeter excess in GRS~1734-292, along with its observed time variability, can be attributed to radio synchrotron emission originating from the corona with the magnetic field $B\approx10$\,G and the size of $R\approx10\,R_{\rm s}$.
The discussion of potential gamma-ray and neutrino fluxes based on the coronal parameters in this paper is reserved for future work. This future analysis will be crucial in establishing a connection between millimeter observations and high-energy astrophysics.
 Nonetheless, it is worth considering that synchrotron and free-free emission from accelerated electrons in $\approx10$\,pc disk winds could also potentially contribute to the observed millimeter excess. To distinguish between these two scenarios, we need to achieve the angular resolution of $0\farcs01$ ($\approx4$\,pc) that can be achieved by the longest ALMA baseline.
Furthermore, it is crucial to validate the light curve using more than two observing epochs in both millimeter bands and X-ray in order to gain a better understanding of the activities taking place in the coronal region.

\begin{acknowledgments}
The authors thank the anonymous referee for carefully reviewing the manuscript.
T.M., S.B., and Y.I. appreciate support from NAOJ ALMA Scientific Research Grant Number 2021-17A.
T.M. is supported by JSPS KAKENHI grant No. 22K14073.
T.M. was supported by the ALMA Japan Research Grant of NAOJ ALMA Project, NAOJ-ALMA-302.
Y.I. is supported by JSPS KAKENHI Grant Number JP18H05458, JP19K14772, and JP22K18277.
This paper makes use of the following ALMA data: ADS/JAO.ALMA 2018.1.00576.S,  ALMA is a partnership of ESO (representing its member states), NSF (USA) and NINS (Japan), together with NRC (Canada), MOST and ASIAA (Taiwan), and KASI (Republic of Korea), in cooperation with the Republic of Chile. 
Some of the ALMA data were retrieved from the JVO portal (\url{http://jvo.nao.ac.jp/portal/}) operated by ADC/NAOJ. Data analysis was in part carried out on the common use data analysis computer system at the Astronomy Data Center, ADC, of the National Astronomical Observatory of Japan. The Joint ALMA Observatory is operated by ESO, AUI/NRAO, and NAOJ.

%
\vspace{5mm}
\facilities{ALMA, VLA}
\software{
NumPy \citep{Harris2020Natur.585..357H},
SciPy \citep{2020SciPy-NMeth},
Matplotlib \citep{Hunter:2007},
Astropy \citep{2022ApJ...935..167A,2018AJ....156..123A,2013A&A...558A..33A},
ALMA Calibration Pipeline,
and CASA \citep{McMullin_2007ASPC..376..127M,CASA2022arXiv221002276T}.
}
\end{acknowledgments}

%




\bibliography{sample631}{}

\begin{thebibliography}{}
\expandafter\ifx\csname natexlab\endcsname\relax\def\natexlab#1{#1}\fi
\providecommand{\url}[1]{\href{#1}{#1}}
\providecommand{\dodoi}[1]{doi:~\href{http://doi.org/#1}{\nolinkurl{#1}}}
\providecommand{\doeprint}[1]{\href{http://ascl.net/#1}{\nolinkurl{http://ascl.net/#1}}}
\providecommand{\doarXiv}[1]{\href{https://arxiv.org/abs/#1}{\nolinkurl{https://arxiv.org/abs/#1}}}

\bibitem[{{Abdollahi} {et~al.}(2022){Abdollahi}, {Acero}, {Baldini}, {Ballet},
  {Bastieri}, {Bellazzini}, {Berenji}, {Berretta}, {Bissaldi}, {Blandford},
  {Bloom}, {Bonino}, {Brill}, {Britto}, {Bruel}, {Burnett}, {Buson}, {Cameron},
  {Caputo}, {Caraveo}, {Castro}, {Chaty}, {Cheung}, {Chiaro}, {Cibrario},
  {Ciprini}, {Coronado-Bl{\'a}zquez}, {Crnogorcevic}, {Cutini}, {D'Ammando},
  {De Gaetano}, {Digel}, {Di Lalla}, {Dirirsa}, {Di Venere}, {Dom{\'\i}nguez},
  {Fallah Ramazani}, {Fegan}, {Ferrara}, {Fiori}, {Fleischhack}, {Franckowiak},
  {Fukazawa}, {Funk}, {Fusco}, {Galanti}, {Gammaldi}, {Gargano}, {Garrappa},
  {Gasparrini}, {Giacchino}, {Giglietto}, {Giordano}, {Giroletti}, {Glanzman},
  {Green}, {Grenier}, {Grondin}, {Guillemot}, {Guiriec}, {Gustafsson},
  {Harding}, {Hays}, {Hewitt}, {Horan}, {Hou}, {J{\'o}hannesson}, {Karwin},
  {Kayanoki}, {Kerr}, {Kuss}, {Landriu}, {Larsson}, {Latronico},
  {Lemoine-Goumard}, {Li}, {Liodakis}, {Longo}, {Loparco}, {Lott}, {Lubrano},
  {Maldera}, {Malyshev}, {Manfreda}, {Mart{\'\i}-Devesa}, {Mazziotta}, {Mereu},
  {Meyer}, {Michelson}, {Mirabal}, {Mitthumsiri}, {Mizuno}, {Moiseev},
  {Monzani}, {Morselli}, {Moskalenko}, {Negro}, {Nuss}, {Omodei}, {Orienti},
  {Orlando}, {Paneque}, {Pei}, {Perkins}, {Persic}, {Pesce-Rollins},
  {Petrosian}, {Pillera}, {Poon}, {Porter}, {Principe}, {Rain{\`o}}, {Rando},
  {Rani}, {Razzano}, {Razzaque}, {Reimer}, {Reimer}, {Reposeur},
  {S{\'a}nchez-Conde}, {Saz Parkinson}, {Scotton}, {Serini}, {Sgr{\`o}},
  {Siskind}, {Smith}, {Spandre}, {Spinelli}, {Sueoka}, {Suson}, {Tajima},
  {Tak}, {Thayer}, {Thompson}, {Torres}, {Troja}, {Valverde}, {Wood}, \&
  {Zaharijas}}]{Abdollahi2022ApJS..260...53A}
{Abdollahi}, S., {Acero}, F., {Baldini}, L., {et~al.} 2022, \apjs, 260, 53,
  \dodoi{10.3847/1538-4365/ac6751}

\bibitem[{{Acciari} {et~al.}(2019){Acciari}, {Ansoldi}, {Antonelli}, {Arbet
  Engels}, {Baack}, {Babi{\'c}}, {Banerjee}, {Barres de Almeida}, {Barrio},
  {Becerra Gonz{\'a}lez}, {Bednarek}, {Bellizzi}, {Bernardini}, {Berti},
  {Besenrieder}, {Bhattacharyya}, {Bigongiari}, {Biland}, {Blanch}, {Bonnoli},
  {Bo{\v{s}}njak}, {Busetto}, {Carosi}, {Ceribella}, {Chai}, {Chilingaryan},
  {Cikota}, {Colak}, {Colin}, {Colombo}, {Contreras}, {Cortina}, {Covino},
  {D'Elia}, {Da Vela}, {Dazzi}, {De Angelis}, {De Lotto}, {Delfino}, {Delgado},
  {Depaoli}, {Di Pierro}, {Di Venere}, {Do Souto Espi{\~n}eira}, {Dominis
  Prester}, {Donini}, {Dorner}, {Doro}, {Elsaesser}, {Fallah Ramazani},
  {Fattorini}, {Ferrara}, {Fidalgo}, {Foffano}, {Fonseca}, {Font}, {Fruck},
  {Fukami}, {Garc{\'\i}a L{\'o}pez}, {Garczarczyk}, {Gasparyan}, {Gaug},
  {Giglietto}, {Giordano}, {Godinovi{\'c}}, {Green}, {Guberman}, {Hadasch},
  {Hahn}, {Herrera}, {Hoang}, {Hrupec}, {H{\"u}tten}, {Inada}, {Inoue},
  {Ishio}, {Iwamura}, {Jouvin}, {Kerszberg}, {Kubo}, {Kushida}, {Lamastra},
  {Lelas}, {Leone}, {Lindfors}, {Lombardi}, {Longo}, {L{\'o}pez},
  {L{\'o}pez-Coto}, {L{\'o}pez-Oramas}, {Loporchio}, {Machado de Oliveira
  Fraga}, {Maggio}, {Majumdar}, {Makariev}, {Mallamaci}, {Maneva}, {Manganaro},
  {Mannheim}, {Maraschi}, {Mariotti}, {Mart{\'\i}nez}, {Mazin},
  {Mi{\'c}anovi{\'c}}, {Miceli}, {Minev}, {Miranda}, {Mirzoyan}, {Molina},
  {Moralejo}, {Morcuende}, {Moreno}, {Moretti}, {Munar-Adrover}, {Neustroev},
  {Nigro}, {Nilsson}, {Ninci}, {Nishijima}, {Noda}, {Nogu{\'e}s}, {Nozaki},
  {Paiano}, {Palacio}, {Palatiello}, {Paneque}, {Paoletti}, {Paredes},
  {Pe{\~n}il}, {Peresano}, {Persic}, {Prada Moroni}, {Prandini}, {Puljak},
  {Rhode}, {Rib{\'o}}, {Rico}, {Righi}, {Rugliancich}, {Saha}, {Sahakyan},
  {Saito}, {Sakurai}, {Satalecka}, {Schmidt}, {Schweizer}, {Sitarek},
  {{\v{S}}nidari{\'c}}, {Sobczynska}, {Somero}, {Stamerra}, {Strom}, {Strzys},
  {Suda}, {Suri{\'c}}, {Takahashi}, {Tavecchio}, {Temnikov}, {Terzi{\'c}},
  {Teshima}, {Torres-Alb{\`a}}, {Tosti}, {Vagelli}, {van Scherpenberg},
  {Vanzo}, {Vazquez Acosta}, {Vigorito}, {Vitale}, {Vovk}, {Will}, {Zari{\'c}},
  {MAGIC Collaboration}, {Fiore}, {Feruglio}, \&
  {Rephaeli}}]{Acciari2019ApJ...883..135A}
{Acciari}, V.~A., {Ansoldi}, S., {Antonelli}, L.~A., {et~al.} 2019, \apj, 883,
  135, \dodoi{10.3847/1538-4357/ab3a51}

\bibitem[{{Ackermann} {et~al.}(2012){Ackermann}, {Ajello}, {Allafort},
  {Baldini}, {Ballet}, {Barbiellini}, {Bastieri}, {Bechtol}, {Bellazzini},
  {Berenji}, {Bloom}, {Bonamente}, {Borgland}, {Bregeon}, {Brigida}, {Bruel},
  {Buehler}, {Buson}, {Caliandro}, {Cameron}, {Caraveo}, {Casandjian},
  {Cavazzuti}, {Cecchi}, {Charles}, {Chekhtman}, {Cheung}, {Chiang}, {Ciprini},
  {Claus}, {Cohen-Tanugi}, {Conrad}, {Cutini}, {D'Ammando}, {de Angelis}, {de
  Palma}, {Dermer}, {Silva}, {Drell}, {Drlica-Wagner}, {Enoto}, {Favuzzi},
  {Fegan}, {Ferrara}, {Fortin}, {Fukazawa}, {Fusco}, {Gargano}, {Gasparrini},
  {Gehrels}, {Germani}, {Giglietto}, {Giommi}, {Giordano}, {Giroletti},
  {Godfrey}, {Grove}, {Guiriec}, {Hadasch}, {Hayashida}, {Hays}, {Hughes},
  {J{\'o}hannesson}, {Johnson}, {Kamae}, {Katagiri}, {Kataoka},
  {Kn{\"o}dlseder}, {Kuss}, {Lande}, {Llena Garde}, {Longo}, {Loparco}, {Lott},
  {Lovellette}, {Lubrano}, {Madejski}, {Mazziotta}, {Michelson}, {Mizuno},
  {Monte}, {Monzani}, {Morselli}, {Moskalenko}, {Murgia}, {Nishino}, {Norris},
  {Nuss}, {Ohno}, {Ohsugi}, {Okumura}, {Orlando}, {Ozaki}, {Paneque},
  {Pesce-Rollins}, {Pierbattista}, {Piron}, {Pivato}, {Porter}, {Rain{\`o}},
  {Rando}, {Razzano}, {Reimer}, {Reimer}, {Ritz}, {Roth}, {Sanchez}, {Sbarra},
  {Sgr{\`o}}, {Siskind}, {Spandre}, {Spinelli}, {Stawarz}, {Strong},
  {Takahashi}, {Takahashi}, {Tanaka}, {Thayer}, {Thompson}, {Tibaldo},
  {Tinivella}, {Torres}, {Tosti}, {Troja}, {Uchiyama}, {Usher},
  {Vandenbroucke}, {Vasileiou}, {Vianello}, {Vitale}, {Waite}, {Winer}, {Wood},
  {Wood}, {Yang}, \& {Zimmer}}]{Ackermann2012ApJ...747..104A}
{Ackermann}, M., {Ajello}, M., {Allafort}, A., {et~al.} 2012, \apj, 747, 104,
  \dodoi{10.1088/0004-637X/747/2/104}

\bibitem[{{Ajello} {et~al.}(2020){Ajello}, {Di Mauro}, {Paliya}, \&
  {Garrappa}}]{Ajello2020ApJ...894...88A}
{Ajello}, M., {Di Mauro}, M., {Paliya}, V.~S., \& {Garrappa}, S. 2020, \apj,
  894, 88, \dodoi{10.3847/1538-4357/ab86a6}

\bibitem[{{Ajello} {et~al.}(2021){Ajello}, {Baldini}, {Ballet}, {Barbiellini},
  {Bastieri}, {Bellazzini}, {Berretta}, {Bissaldi}, {Blandford}, {Bloom},
  {Bonino}, {Bruel}, {Buson}, {Cameron}, {Caprioli}, {Caputo}, {Cavazzuti},
  {Chartas}, {Chen}, {Cheung}, {Chiaro}, {Costantin}, {Cutini}, {D'Ammando},
  {de la Torre Luque}, {de Palma}, {Desai}, {Diesing}, {Di Lalla}, {Dirirsa},
  {Di Venere}, {Dom{\'\i}nguez}, {Fegan}, {Franckowiak}, {Fukazawa}, {Funk},
  {Fusco}, {Gargano}, {Gasparrini}, {Giglietto}, {Giordano}, {Giroletti},
  {Green}, {Grenier}, {Guiriec}, {Hartmann}, {Horan}, {J{\'o}hannesson},
  {Karwin}, {Kerr}, {Kova{\v{c}}evi{\'c}}, {Kuss}, {Larsson}, {Latronico},
  {Lemoine-Goumard}, {Li}, {Liodakis}, {Longo}, {Loparco}, {Lovellette},
  {Lubrano}, {Maldera}, {Manfreda}, {Marchesi}, {Marcotulli},
  {Mart{\'\i}-Devesa}, {Mazziotta}, {Mereu}, {Michelson}, {Mizuno}, {Monzani},
  {Morselli}, {Moskalenko}, {Negro}, {Omodei}, {Orienti}, {Orlando}, {Paliya},
  {Paneque}, {Pei}, {Persic}, {Pesce-Rollins}, {Porter}, {Principe}, {Racusin},
  {Rain{\`o}}, {Rando}, {Rani}, {Razzano}, {Reimer}, {Reimer}, {Saz Parkinson},
  {Serini}, {Sgr{\`o}}, {Siskind}, {Spandre}, {Spinelli}, {Suson}, {Tak},
  {Torres}, {Troja}, {Wood}, {Zaharijas}, \&
  {Zrake}}]{Ajello2021ApJ...921..144A}
{Ajello}, M., {Baldini}, L., {Ballet}, J., {et~al.} 2021, \apj, 921, 144,
  \dodoi{10.3847/1538-4357/ac1bb2}

\bibitem[{{Astropy Collaboration} {et~al.}(2013){Astropy Collaboration},
  {Robitaille}, {Tollerud}, {Greenfield}, {Droettboom}, {Bray}, {Aldcroft},
  {Davis}, {Ginsburg}, {Price-Whelan}, {Kerzendorf}, {Conley}, {Crighton},
  {Barbary}, {Muna}, {Ferguson}, {Grollier}, {Parikh}, {Nair}, {Unther},
  {Deil}, {Woillez}, {Conseil}, {Kramer}, {Turner}, {Singer}, {Fox}, {Weaver},
  {Zabalza}, {Edwards}, {Azalee Bostroem}, {Burke}, {Casey}, {Crawford},
  {Dencheva}, {Ely}, {Jenness}, {Labrie}, {Lim}, {Pierfederici}, {Pontzen},
  {Ptak}, {Refsdal}, {Servillat}, \& {Streicher}}]{2013A&A...558A..33A}
{Astropy Collaboration}, {Robitaille}, T.~P., {Tollerud}, E.~J., {et~al.} 2013,
  \aap, 558, A33, \dodoi{10.1051/0004-6361/201322068}

\bibitem[{{Astropy Collaboration} {et~al.}(2018){Astropy Collaboration},
  {Price-Whelan}, {Sip{\H{o}}cz}, {G{\"u}nther}, {Lim}, {Crawford}, {Conseil},
  {Shupe}, {Craig}, {Dencheva}, {Ginsburg}, {VanderPlas}, {Bradley},
  {P{\'e}rez-Su{\'a}rez}, {de Val-Borro}, {Aldcroft}, {Cruz}, {Robitaille},
  {Tollerud}, {Ardelean}, {Babej}, {Bach}, {Bachetti}, {Bakanov}, {Bamford},
  {Barentsen}, {Barmby}, {Baumbach}, {Berry}, {Biscani}, {Boquien}, {Bostroem},
  {Bouma}, {Brammer}, {Bray}, {Breytenbach}, {Buddelmeijer}, {Burke},
  {Calderone}, {Cano Rodr{\'\i}guez}, {Cara}, {Cardoso}, {Cheedella}, {Copin},
  {Corrales}, {Crichton}, {D'Avella}, {Deil}, {Depagne}, {Dietrich}, {Donath},
  {Droettboom}, {Earl}, {Erben}, {Fabbro}, {Ferreira}, {Finethy}, {Fox},
  {Garrison}, {Gibbons}, {Goldstein}, {Gommers}, {Greco}, {Greenfield},
  {Groener}, {Grollier}, {Hagen}, {Hirst}, {Homeier}, {Horton}, {Hosseinzadeh},
  {Hu}, {Hunkeler}, {Ivezi{\'c}}, {Jain}, {Jenness}, {Kanarek}, {Kendrew},
  {Kern}, {Kerzendorf}, {Khvalko}, {King}, {Kirkby}, {Kulkarni}, {Kumar},
  {Lee}, {Lenz}, {Littlefair}, {Ma}, {Macleod}, {Mastropietro}, {McCully},
  {Montagnac}, {Morris}, {Mueller}, {Mumford}, {Muna}, {Murphy}, {Nelson},
  {Nguyen}, {Ninan}, {N{\"o}the}, {Ogaz}, {Oh}, {Parejko}, {Parley}, {Pascual},
  {Patil}, {Patil}, {Plunkett}, {Prochaska}, {Rastogi}, {Reddy Janga},
  {Sabater}, {Sakurikar}, {Seifert}, {Sherbert}, {Sherwood-Taylor}, {Shih},
  {Sick}, {Silbiger}, {Singanamalla}, {Singer}, {Sladen}, {Sooley},
  {Sornarajah}, {Streicher}, {Teuben}, {Thomas}, {Tremblay}, {Turner},
  {Terr{\'o}n}, {van Kerkwijk}, {de la Vega}, {Watkins}, {Weaver}, {Whitmore},
  {Woillez}, {Zabalza}, \& {Astropy Contributors}}]{2018AJ....156..123A}
{Astropy Collaboration}, {Price-Whelan}, A.~M., {Sip{\H{o}}cz}, B.~M., {et~al.}
  2018, \aj, 156, 123, \dodoi{10.3847/1538-3881/aabc4f}

\bibitem[{{Astropy Collaboration} {et~al.}(2022){Astropy Collaboration},
  {Price-Whelan}, {Lim}, {Earl}, {Starkman}, {Bradley}, {Shupe}, {Patil},
  {Corrales}, {Brasseur}, {N{\"o}the}, {Donath}, {Tollerud}, {Morris},
  {Ginsburg}, {Vaher}, {Weaver}, {Tocknell}, {Jamieson}, {van Kerkwijk},
  {Robitaille}, {Merry}, {Bachetti}, {G{\"u}nther}, {Aldcroft},
  {Alvarado-Montes}, {Archibald}, {B{\'o}di}, {Bapat}, {Barentsen},
  {Baz{\'a}n}, {Biswas}, {Boquien}, {Burke}, {Cara}, {Cara}, {Conroy},
  {Conseil}, {Craig}, {Cross}, {Cruz}, {D'Eugenio}, {Dencheva}, {Devillepoix},
  {Dietrich}, {Eigenbrot}, {Erben}, {Ferreira}, {Foreman-Mackey}, {Fox},
  {Freij}, {Garg}, {Geda}, {Glattly}, {Gondhalekar}, {Gordon}, {Grant},
  {Greenfield}, {Groener}, {Guest}, {Gurovich}, {Handberg}, {Hart},
  {Hatfield-Dodds}, {Homeier}, {Hosseinzadeh}, {Jenness}, {Jones}, {Joseph},
  {Kalmbach}, {Karamehmetoglu}, {Ka{\l}uszy{\'n}ski}, {Kelley}, {Kern},
  {Kerzendorf}, {Koch}, {Kulumani}, {Lee}, {Ly}, {Ma}, {MacBride}, {Maljaars},
  {Muna}, {Murphy}, {Norman}, {O'Steen}, {Oman}, {Pacifici}, {Pascual},
  {Pascual-Granado}, {Patil}, {Perren}, {Pickering}, {Rastogi}, {Roulston},
  {Ryan}, {Rykoff}, {Sabater}, {Sakurikar}, {Salgado}, {Sanghi}, {Saunders},
  {Savchenko}, {Schwardt}, {Seifert-Eckert}, {Shih}, {Jain}, {Shukla}, {Sick},
  {Simpson}, {Singanamalla}, {Singer}, {Singhal}, {Sinha}, {Sip{\H{o}}cz},
  {Spitler}, {Stansby}, {Streicher}, {{\v{S}}umak}, {Swinbank}, {Taranu},
  {Tewary}, {Tremblay}, {Val-Borro}, {Van Kooten}, {Vasovi{\'c}}, {Verma}, {de
  Miranda Cardoso}, {Williams}, {Wilson}, {Winkel}, {Wood-Vasey}, {Xue},
  {Yoachim}, {Zhang}, {Zonca}, \& {Astropy Project
  Contributors}}]{2022ApJ...935..167A}
{Astropy Collaboration}, {Price-Whelan}, A.~M., {Lim}, P.~L., {et~al.} 2022,
  \apj, 935, 167, \dodoi{10.3847/1538-4357/ac7c74}

\bibitem[{{Baumgartner} {et~al.}(2013){Baumgartner}, {Tueller}, {Markwardt},
  {Skinner}, {Barthelmy}, {Mushotzky}, {Evans}, \&
  {Gehrels}}]{Baumgartner2013ApJS..207...19B}
{Baumgartner}, W.~H., {Tueller}, J., {Markwardt}, C.~B., {et~al.} 2013, \apjs,
  207, 19, \dodoi{10.1088/0067-0049/207/2/19}

\bibitem[{{B{\^\i}rzan} {et~al.}(2008){B{\^\i}rzan}, {McNamara}, {Nulsen},
  {Carilli}, \& {Wise}}]{Birzan2008ApJ...686..859B}
{B{\^\i}rzan}, L., {McNamara}, B.~R., {Nulsen}, P.~E.~J., {Carilli}, C.~L., \&
  {Wise}, M.~W. 2008, \apj, 686, 859, \dodoi{10.1086/591416}

\bibitem[{{Chaty} {et~al.}(2011){Chaty}, {Dubus}, \&
  {Raichoor}}]{Chaty2011A&A...529A...3C}
{Chaty}, S., {Dubus}, G., \& {Raichoor}, A. 2011, \aap, 529, A3,
  \dodoi{10.1051/0004-6361/201015589}

\bibitem[{{Di Cocco} {et~al.}(2004){Di Cocco}, {Foschini}, {Grandi},
  {Malaguti}, {Castro-Tirado}, {Chaty}, {Dean}, {Gehrels}, {Grenier},
  {Hermsen}, {Kuiper}, {Lund}, \& {Mirabel}}]{DiCocco2004A&A...425...89D}
{Di Cocco}, G., {Foschini}, L., {Grandi}, P., {et~al.} 2004, \aap, 425, 89,
  \dodoi{10.1051/0004-6361:20040484}

\bibitem[{{Fabian} {et~al.}(2015){Fabian}, {Lohfink}, {Kara}, {Parker},
  {Vasudevan}, \& {Reynolds}}]{Fabian2015MNRAS.451.4375F}
{Fabian}, A.~C., {Lohfink}, A., {Kara}, E., {et~al.} 2015, \mnras, 451, 4375,
  \dodoi{10.1093/mnras/stv1218}

\bibitem[{{Foreman-Mackey} {et~al.}(2013){Foreman-Mackey}, {Hogg}, {Lang}, \&
  {Goodman}}]{Foreman-Mackey2013PASP..125..306F}
{Foreman-Mackey}, D., {Hogg}, D.~W., {Lang}, D., \& {Goodman}, J. 2013, \pasp,
  125, 306, \dodoi{10.1086/670067}

\bibitem[{{Fukazawa} {et~al.}(2022){Fukazawa}, {Matake}, {Kayanoki}, {Inoue},
  \& {Finke}}]{Fukazawa2022ApJ...931..138F}
{Fukazawa}, Y., {Matake}, H., {Kayanoki}, T., {Inoue}, Y., \& {Finke}, J. 2022,
  \apj, 931, 138, \dodoi{10.3847/1538-4357/ac6acb}

\bibitem[{{Galeev} {et~al.}(1979){Galeev}, {Rosner}, \&
  {Vaiana}}]{Galeev1979ApJ...229..318G}
{Galeev}, A.~A., {Rosner}, R., \& {Vaiana}, G.~S. 1979, \apj, 229, 318,
  \dodoi{10.1086/156957}

\bibitem[{{Harris} {et~al.}(2020){Harris}, {Millman}, {van der Walt},
  {Gommers}, {Virtanen}, {Cournapeau}, {Wieser}, {Taylor}, {Berg}, {Smith},
  {Kern}, {Picus}, {Hoyer}, {van Kerkwijk}, {Brett}, {Haldane}, {del R{\'\i}o},
  {Wiebe}, {Peterson}, {G{\'e}rard-Marchant}, {Sheppard}, {Reddy}, {Weckesser},
  {Abbasi}, {Gohlke}, \& {Oliphant}}]{Harris2020Natur.585..357H}
{Harris}, C.~R., {Millman}, K.~J., {van der Walt}, S.~J., {et~al.} 2020, \nat,
  585, 357, \dodoi{10.1038/s41586-020-2649-2}

\bibitem[{{Hayashida} {et~al.}(2013){Hayashida}, {Stawarz}, {Cheung},
  {Bechtol}, {Madejski}, {Ajello}, {Massaro}, {Moskalenko}, {Strong}, \&
  {Tibaldo}}]{Hayashida2013ApJ...779..131H}
{Hayashida}, M., {Stawarz}, {\L}., {Cheung}, C.~C., {et~al.} 2013, \apj, 779,
  131, \dodoi{10.1088/0004-637X/779/2/131}

\bibitem[{{Hills}(1975)}]{Hills1975Natur.254..295H}
{Hills}, J.~G. 1975, \nat, 254, 295, \dodoi{10.1038/254295a0}

\bibitem[{Hunter(2007)}]{Hunter:2007}
Hunter, J.~D. 2007, Computing in Science \& Engineering, 9, 90,
  \dodoi{10.1109/MCSE.2007.55}

\bibitem[{{IceCube Collaboration} {et~al.}(2022){IceCube Collaboration},
  {Abbasi}, {Ackermann}, {Adams}, {Aguilar}, {Ahlers}, {Ahrens}, {Alameddine},
  {Alispach}, {Alves}, {Amin}, {Andeen}, {Anderson}, {Anton}, {Arg{\"u}elles},
  {Ashida}, {Axani}, {Bai}, {Balagopal}, {Barbano}, {Barwick}, {Bastian},
  {Basu}, {Baur}, {Bay}, {Beatty}, {Becker}, {Becker Tjus}, {Bellenghi},
  {Benzvi}, {Berley}, {Bernardini}, {Besson}, {Binder}, {Bindig}, {Blaufuss},
  {Blot}, {Boddenberg}, {Bontempo}, {Borowka}, {B{\"o}ser}, {Botner},
  {B{\"o}ttcher}, {Bourbeau}, {Bradascio}, {Braun}, {Brinson}, {Bron},
  {Brostean-Kaiser}, {Browne}, {Burgman}, {Burley}, {Busse}, {Campana},
  {Carnie-Bronca}, {Chen}, {Chen}, {Chirkin}, {Choi}, {Clark}, {Clark},
  {Classen}, {Coleman}, {Collin}, {Conrad}, {Coppin}, {Correa}, {Cowen},
  {Cross}, {Dappen}, {Dave}, {de Clercq}, {Delaunay}, {Delgado L{\'o}pez},
  {Dembinski}, {Deoskar}, {Desai}, {Desiati}, {de Vries}, {de Wasseige}, {de
  With}, {Deyoung}, {Diaz}, {D{\'\i}az-V{\'e}lez}, {Dittmer}, {Dujmovic},
  {Dunkman}, {Duvernois}, {Dvorak}, {Ehrhardt}, {Eller}, {Engel}, {Erpenbeck},
  {Evans}, {Evenson}, {Fan}, {Fazely}, {Fedynitch}, {Feigl}, {Fiedlschuster},
  {Fienberg}, {Filimonov}, {Finley}, {Fischer}, {Fox}, {Franckowiak},
  {Friedman}, {Fritz}, {F{\"u}rst}, {Gaisser}, {Gallagher}, {Ganster},
  {Garcia}, {Garrappa}, {Gerhardt}, {Ghadimi}, {Glaser}, {Glauch},
  {Gl{\"u}senkamp}, {Goldschmidt}, {Gonzalez}, {Goswami}, {Grant},
  {Gr{\'e}goire}, {Griswold}, {G{\"u}nther}, {Gutjahr}, {Haack}, {Hallgren},
  {Halliday}, {Halve}, {Halzen}, {Hanson}, {Hardin}, {Harnisch}, {Haungs},
  {Hebecker}, {Helbing}, {Henningsen}, {Hettinger}, {Hickford}, {Hignight},
  {Hill}, {Hill}, {Hoffman}, {Hoffmann}, {Hokanson-Fasig}, {Hoshina}, {Huang},
  {Huber}, {Huber}, {Hultqvist}, {H{\"u}nnefeld}, {Hussain}, {Hymon}, {in},
  {Iovine}, {Ishihara}, {Jansson}, {Japaridze}, {Jeong}, {Jin}, {Jones},
  {Kang}, {Kang}, {Kang}, {Kappes}, {Kappesser}, {Kardum}, {Karg}, {Karl},
  {Karle}, {Katz}, {Kauer}, {Kellermann}, {Kelley}, {Kheirandish}, {Kin},
  {Kintscher}, {Kiryluk}, {Klein}, {Koirala}, {Kolanoski}, {Kontrimas},
  {K{\"o}pke}, {Kopper}, {Kopper}, {Koskinen}, {Koundal}, {Kovacevich},
  {Kowalski}, {Kozynets}, {Kun}, {Kurahashi}, {Lad}, {Lagunas Gualda},
  {Lanfranchi}, {Larson}, {Lauber}, {Lazar}, {Lee}, {Leonard},
  {Leszczy{\'n}ska}, {Li}, {Lincetto}, {Liu}, {Liubarska}, {Lohfink}, {Lozano
  Mariscal}, {Lu}, {Lucarelli}, {Ludwig}, {Luszczak}, {Lyu}, {Ma}, {Madsen},
  {Mahn}, {Makino}, {Mancina}, {Mari{\c{s}}}, {Martinez-Soler}, {Maruyama},
  {Mase}, {McElroy}, {McNally}, {Mead}, {Meagher}, {Mechbal}, {Medina},
  {Meier}, {Meighen-Berger}, {Micallef}, {Mockler}, {Montaruli}, {Moore},
  {Morse}, {Moulai}, {Naab}, {Nagai}, {Nahnhauer}, {Naumann}, {Necker},
  {Nguyen}, {Niederhausen}, {Nisa}, {Nowicki}, {Nygren}, {Obertack},
  {Pollmann}, {Oehler}, {Oeyen}, {Olivas}, {O'Sullivan}, {Pandya}, {Pankova},
  {Park}, {Parker}, {Paudel}, {Paul}, {P{\'e}rez de Los Heros}, {Peters},
  {Peterson}, {Philippen}, {Pieper}, {Pittermann}, {Pizzuto}, {Plum},
  {Popovych}, {Porcelli}, {Prado Rodriguez}, {Price}, {Pries}, {Przybylski},
  {Rack-Helleis}, {Raissi}, {Rameez}, {Rawlins}, {Rea}, {Rehman},
  {Reichherzer}, {Reimann}, {Renzi}, {Resconi}, {Reusch}, {Rhode}, {Richman},
  {Riedel}, {Roberts}, {Robertson}, {Roellinghoff}, {Rongen}, {Rott}, {Ruhe},
  {Ryckbosch}, {Rysewyk Cantu}, {Safa}, {Saffer}, {Sanchez Herrera},
  {Sandrock}, {Sandroos}, {Santander}, {Sarkar}, {Sarkar}, {Satalecka},
  {Schaufel}, {Schieler}, {Schindler}, {Schmidt}, {Schneider}, {Schneider},
  {Schr{\"o}der}, {Schumacher}, {Schwefer}, {Sclafani}, {Seckel}, {Seunarine},
  {Sharma}, {Shefali}, {Silva}, {Skrzypek}, {Smithers}, {Snihur},
  {Soedingrekso}, {Soldin}, {Spannfellner}, {Spiczak}, {Spiering},
  {Stachurska}, {Stamatikos}, {Stanev}, {Stein}, {Stettner}, {Steuer},
  {Stezelberger}, {Stokstad}, {St{\"u}rwald}, {Stuttard}, {Sullivan},
  {Taboada}, {Ter-Antonyan}, {Tilav}, {Tischbein}, {Tollefson}, {T{\"o}nnis},
  {Toscano}, {Tosi}, {Trettin}, {Tselengidou}, {Tung}, {Turcati}, {Turcotte},
  {Turley}, {Twagirayezu}, {Ty}, {Unland Elorrieta}, {Valtonen-Mattila},
  {Vandenbroucke}, {van Eijndhoven}, {Vannerom}, {van Santen}, {Verpoest},
  {Walck}, {Watson}, {Weaver}, {Weigel}, {Weindl}, {Weiss}, {Weldert}, {Wendt},
  {Werthebach}, {Weyrauch}, {Whitehorn}, {Wiebusch}, {Williams}, {Wolf},
  {Woschnagg}, {Wrede}, {Wulff}, {Xu}, {Yanez}, {Yoshida}, {Yu}, {Yuan},
  {Zhangan}, \& {Zhelnin}}]{IceCube2022Sci...378..538I}
{IceCube Collaboration}, {Abbasi}, R., {Ackermann}, M., {et~al.} 2022, Science,
  378, 538, \dodoi{10.1126/science.abg3395}

\bibitem[{{Inoue}(2011)}]{Inoue2011ApJ...733...66I}
{Inoue}, Y. 2011, \apj, 733, 66, \dodoi{10.1088/0004-637X/733/1/66}

\bibitem[{{Inoue} \& {Doi}(2014)}]{Inoue2014PASJ...66L...8I}
{Inoue}, Y., \& {Doi}, A. 2014, \pasj, 66, L8, \dodoi{10.1093/pasj/psu079}

\bibitem[{{Inoue} \& {Doi}(2018)}]{Inoue2018ApJ...869..114I}
---. 2018, \apj, 869, 114, \dodoi{10.3847/1538-4357/aaeb95}

\bibitem[{{Inoue} {et~al.}(2020){Inoue}, {Khangulyan}, \&
  {Doi}}]{Inoue2020ApJ...891L..33I}
{Inoue}, Y., {Khangulyan}, D., \& {Doi}, A. 2020, \apjl, 891, L33,
  \dodoi{10.3847/2041-8213/ab7661}

\bibitem[{{Inoue} {et~al.}(2019){Inoue}, {Khangulyan}, {Inoue}, \&
  {Doi}}]{Inoue2019ApJ...880...40I}
{Inoue}, Y., {Khangulyan}, D., {Inoue}, S., \& {Doi}, A. 2019, \apj, 880, 40,
  \dodoi{10.3847/1538-4357/ab2715}

\bibitem[{{Inoue} {et~al.}(2024){Inoue}, {Takasao}, \&
  {Khangulyan}}]{Inoue2024arXiv240107580I}
{Inoue}, Y., {Takasao}, S., \& {Khangulyan}, D. 2024, arXiv e-prints,
  arXiv:2401.07580, \dodoi{10.48550/arXiv.2401.07580}

\bibitem[{{Kawamuro} {et~al.}(2022){Kawamuro}, {Ricci}, {Imanishi},
  {Mushotzky}, {Izumi}, {Ricci}, {Bauer}, {Koss}, {Trakhtenbrot}, {Ichikawa},
  {Rojas}, {Smith}, {Shimizu}, {Oh}, {den Brok}, {Baba}, {Balokovi{\'c}},
  {Chang}, {Kakkad}, {Pfeifle}, {Privon}, {Temple}, {Ueda}, {Harrison},
  {Powell}, {Stern}, {Urry}, \& {Sanders}}]{Kawamuro2022ApJ...938...87K}
{Kawamuro}, T., {Ricci}, C., {Imanishi}, M., {et~al.} 2022, \apj, 938, 87,
  \dodoi{10.3847/1538-4357/ac8794}

\bibitem[{{Kawamuro} {et~al.}(2023){Kawamuro}, {Ricci}, {Mushotzky},
  {Imanishi}, {Bauer}, {Ricci}, {Koss}, {Privon}, {Trakhtenbrot}, {Izumi},
  {Ichikawa}, {Rojas}, {Smith}, {Shimizu}, {Oh}, {den Brok}, {Baba},
  {Balokovi{\'c}}, {Chang}, {Kakkad}, {Pfeifle}, {Temple}, {Ueda}, {Harrison},
  {Powell}, {Stern}, {Urry}, \& {Sanders}}]{Kawamuro2023ApJS..269...24K}
{Kawamuro}, T., {Ricci}, C., {Mushotzky}, R.~F., {et~al.} 2023, \apjs, 269, 24,
  \dodoi{10.3847/1538-4365/acf467}

\bibitem[{{Khachikian} \& {Weedman}(1974)}]{Khachikian1974ApJ...192..581K}
{Khachikian}, E.~Y., \& {Weedman}, D.~W. 1974, \apj, 192, 581,
  \dodoi{10.1086/153093}

\bibitem[{{Laha} {et~al.}(2021){Laha}, {Reynolds}, {Reeves}, {Kriss},
  {Guainazzi}, {Smith}, {Veilleux}, \& {Proga}}]{Laha2021NatAs...5...13L}
{Laha}, S., {Reynolds}, C.~S., {Reeves}, J., {et~al.} 2021, Nature Astronomy,
  5, 13, \dodoi{10.1038/s41550-020-01255-2}

\bibitem[{{Laor} \& {Behar}(2008)}]{Laor2008MNRAS.390..847L}
{Laor}, A., \& {Behar}, E. 2008, \mnras, 390, 847,
  \dodoi{10.1111/j.1365-2966.2008.13806.x}

\bibitem[{{Lenain} {et~al.}(2010){Lenain}, {Ricci}, {T{\"u}rler}, {Dorner}, \&
  {Walter}}]{Lenain2010A&A...524A..72L}
{Lenain}, J.~P., {Ricci}, C., {T{\"u}rler}, M., {Dorner}, D., \& {Walter}, R.
  2010, \aap, 524, A72, \dodoi{10.1051/0004-6361/201015644}

\bibitem[{{Liska} {et~al.}(2022){Liska}, {Musoke}, {Tchekhovskoy}, {Porth}, \&
  {Beloborodov}}]{Liska2022ApJ...935L...1L}
{Liska}, M.~T.~P., {Musoke}, G., {Tchekhovskoy}, A., {Porth}, O., \&
  {Beloborodov}, A.~M. 2022, \apjl, 935, L1, \dodoi{10.3847/2041-8213/ac84db}

\bibitem[{{Luo} {et~al.}(2017){Luo}, {Brandt}, {Xue}, {Lehmer}, {Alexander},
  {Bauer}, {Vito}, {Yang}, {Basu-Zych}, {Comastri}, {Gilli}, {Gu},
  {Hornschemeier}, {Koekemoer}, {Liu}, {Mainieri}, {Paolillo}, {Ranalli},
  {Rosati}, {Schneider}, {Shemmer}, {Smail}, {Sun}, {Tozzi}, {Vignali}, \&
  {Wang}}]{Luo2017ApJS..228....2L}
{Luo}, B., {Brandt}, W.~N., {Xue}, Y.~Q., {et~al.} 2017, \apjs, 228, 2,
  \dodoi{10.3847/1538-4365/228/1/2}

\bibitem[{{Marconi} \& {Hunt}(2003)}]{Marconi2003ApJ...589L..21M}
{Marconi}, A., \& {Hunt}, L.~K. 2003, \apjl, 589, L21, \dodoi{10.1086/375804}

\bibitem[{{Marconi} {et~al.}(2004){Marconi}, {Risaliti}, {Gilli}, {Hunt},
  {Maiolino}, \& {Salvati}}]{Marconi2004MNRAS.351..169M}
{Marconi}, A., {Risaliti}, G., {Gilli}, R., {et~al.} 2004, \mnras, 351, 169,
  \dodoi{10.1111/j.1365-2966.2004.07765.x}

\bibitem[{{Marcotulli} {et~al.}(2020){Marcotulli}, {Di Mauro}, \&
  {Ajello}}]{Marcotulli2020ApJ...896....6M}
{Marcotulli}, L., {Di Mauro}, M., \& {Ajello}, M. 2020, \apj, 896, 6,
  \dodoi{10.3847/1538-4357/ab8cbd}

\bibitem[{{Marti} {et~al.}(1998){Marti}, {Mirabel}, {Chaty}, \&
  {Rodriguez}}]{Marti1998A&A...330...72M}
{Marti}, J., {Mirabel}, I.~F., {Chaty}, S., \& {Rodriguez}, L.~F. 1998, \aap,
  330, 72, \dodoi{10.48550/arXiv.astro-ph/9710136}

\bibitem[{{McMullin} {et~al.}(2007){McMullin}, {Waters}, {Schiebel}, {Young},
  \& {Golap}}]{McMullin_2007ASPC..376..127M}
{McMullin}, J.~P., {Waters}, B., {Schiebel}, D., {Young}, W., \& {Golap}, K.
  2007, in Astronomical Society of the Pacific Conference Series, Vol. 376,
  Astronomical Data Analysis Software and Systems XVI, ed. R.~A. {Shaw},
  F.~{Hill}, \& D.~J. {Bell}, 127

\bibitem[{{Michiyama} {et~al.}(2023){Michiyama}, {Inoue}, \&
  {Doi}}]{Michiyama2023PASJ...75..874M}
{Michiyama}, T., {Inoue}, Y., \& {Doi}, A. 2023, \pasj, 75, 874,
  \dodoi{10.1093/pasj/psad044}

\bibitem[{{Michiyama} {et~al.}(2022){Michiyama}, {Inoue}, {Doi}, \&
  {Khangulyan}}]{Michiyama2022ApJ...936L...1M}
{Michiyama}, T., {Inoue}, Y., {Doi}, A., \& {Khangulyan}, D. 2022, \apjl, 936,
  L1, \dodoi{10.3847/2041-8213/ac8935}

\bibitem[{{Molina} {et~al.}(2019){Molina}, {Malizia}, {Bassani}, {Ursini},
  {Bazzano}, \& {Ubertini}}]{Molina2019MNRAS.484.2735M}
{Molina}, M., {Malizia}, A., {Bassani}, L., {et~al.} 2019, \mnras, 484, 2735,
  \dodoi{10.1093/mnras/stz156}

\bibitem[{{Nims} {et~al.}(2015){Nims}, {Quataert}, \&
  {Faucher-Gigu{\`e}re}}]{Nims2015MNRAS.447.3612N}
{Nims}, J., {Quataert}, E., \& {Faucher-Gigu{\`e}re}, C.-A. 2015, \mnras, 447,
  3612, \dodoi{10.1093/mnras/stu2648}

\bibitem[{{Oh} {et~al.}(2018){Oh}, {Koss}, {Markwardt}, {Schawinski},
  {Baumgartner}, {Barthelmy}, {Cenko}, {Gehrels}, {Mushotzky}, {Petulante},
  {Ricci}, {Lien}, \& {Trakhtenbrot}}]{Oh2018ApJS..235....4O}
{Oh}, K., {Koss}, M., {Markwardt}, C.~B., {et~al.} 2018, \apjs, 235, 4,
  \dodoi{10.3847/1538-4365/aaa7fd}

\bibitem[{{Pal} {et~al.}(2023){Pal}, {Anju}, {Sreehari}, {Rameshan}, {Stalin},
  {Ricci}, \& {Marchesi}}]{Pal2023arXiv231018196P}
{Pal}, I., {Anju}, A., {Sreehari}, H., {et~al.} 2023, arXiv e-prints,
  arXiv:2310.18196, \dodoi{10.48550/arXiv.2310.18196}

\bibitem[{{Parker}(1955)}]{Parker1955ApJ...121..491P}
{Parker}, E.~N. 1955, \apj, 121, 491, \dodoi{10.1086/146010}

\bibitem[{{Parker}(1966)}]{Parker1966ApJ...145..811P}
---. 1966, \apj, 145, 811, \dodoi{10.1086/148828}

\bibitem[{{Peretti} {et~al.}(2023){Peretti}, {Peron}, {Tombesi}, {Lamastra},
  {Ahlers}, \& {Saturni}}]{Peretti2023arXiv230303298P}
{Peretti}, E., {Peron}, G., {Tombesi}, F., {et~al.} 2023, arXiv e-prints,
  arXiv:2303.03298, \dodoi{10.48550/arXiv.2303.03298}

\bibitem[{{Ricci} {et~al.}(2018){Ricci}, {Ho}, {Fabian}, {Trakhtenbrot},
  {Koss}, {Ueda}, {Lohfink}, {Shimizu}, {Bauer}, {Mushotzky}, {Schawinski},
  {Paltani}, {Lamperti}, {Treister}, \& {Oh}}]{Ricci2018MNRAS.480.1819R}
{Ricci}, C., {Ho}, L.~C., {Fabian}, A.~C., {et~al.} 2018, \mnras, 480, 1819,
  \dodoi{10.1093/mnras/sty1879}

\bibitem[{{Ricci} {et~al.}(2023){Ricci}, {Chang}, {Kawamuro}, {Privon},
  {Mushotzky}, {Trakhtenbrot}, {Laor}, {Koss}, {Smith}, {Gupta}, {Dimopoulos},
  {Aalto}, \& {Ros}}]{Ricci2023ApJ...952L..28R}
{Ricci}, C., {Chang}, C.-S., {Kawamuro}, T., {et~al.} 2023, \apjl, 952, L28,
  \dodoi{10.3847/2041-8213/acda27}

\bibitem[{{Sazonov} {et~al.}(2004){Sazonov}, {Revnivtsev}, {Lutovinov},
  {Sunyaev}, \& {Grebenev}}]{Sazonov2004A&A...421L..21S}
{Sazonov}, S.~Y., {Revnivtsev}, M.~G., {Lutovinov}, A.~A., {Sunyaev}, R.~A., \&
  {Grebenev}, S.~A. 2004, \aap, 421, L21, \dodoi{10.1051/0004-6361:20040179}

\bibitem[{{Seyfert}(1943)}]{Seyfert1943ApJ....97...28S}
{Seyfert}, C.~K. 1943, \apj, 97, 28, \dodoi{10.1086/144488}

\bibitem[{{Shimizu} {et~al.}(2016){Shimizu}, {Mel{\'e}ndez}, {Mushotzky},
  {Koss}, {Barger}, \& {Cowie}}]{Shimizu2016MNRAS.456.3335S}
{Shimizu}, T.~T., {Mel{\'e}ndez}, M., {Mushotzky}, R.~F., {et~al.} 2016,
  \mnras, 456, 3335, \dodoi{10.1093/mnras/stv2828}

\bibitem[{{Sunyaev}(1990)}]{Sunyaev1990IAUC.5123....2S}
{Sunyaev}, R. 1990, \iaucirc, 5123, 2

\bibitem[{{Takasao} {et~al.}(2018){Takasao}, {Tomida}, {Iwasaki}, \&
  {Suzuki}}]{Takasao2018ApJ...857....4T}
{Takasao}, S., {Tomida}, K., {Iwasaki}, K., \& {Suzuki}, T.~K. 2018, \apj, 857,
  4, \dodoi{10.3847/1538-4357/aab5b3}

\bibitem[{{THE CASA TEAM} {et~al.}(2022){THE CASA TEAM}, {Bean}, {Bhatnagar},
  {Castro}, {Donovan Meyer}, {Emonts}, {Garcia}, {Garwood}, {Golap}, {Gonzalez
  Villalba}, {Harris}, {Hayashi}, {Hoskins}, {Hsieh}, {Jagannathan},
  {Kawasaki}, {Keimpema}, {Kettenis}, {Lopez}, {Marvil}, {Masters},
  {McNichols}, {Mehringer}, {Miel}, {Moellenbrock}, {Montesino}, {Nakazato},
  {Ott}, {Petry}, {Pokorny}, {Raba}, {Rau}, {Schiebel}, {Schweighart},
  {Sekhar}, {Shimada}, {Small}, {Steeb}, {Sugimoto}, {Suoranta}, {Tsutsumi},
  {van Bemmel}, {Verkouter}, {Wells}, {Xiong}, {Szomoru}, {Griffith},
  {Glendenning}, \& {Kern}}]{CASA2022arXiv221002276T}
{THE CASA TEAM}, {Bean}, B., {Bhatnagar}, S., {et~al.} 2022, arXiv e-prints,
  arXiv:2210.02276.
\newblock \doarXiv{2210.02276}

\bibitem[{{Tortosa} {et~al.}(2017){Tortosa}, {Marinucci}, {Matt}, {Bianchi},
  {La Franca}, {Ballantyne}, {Boorman}, {Fabian}, {Farrah}, {Fuerst}, {Gandhi},
  {Harrison}, {Koss}, {Ricci}, {Stern}, {Ursini}, \&
  {Walton}}]{Tortosa2017MNRAS.466.4193T}
{Tortosa}, A., {Marinucci}, A., {Matt}, G., {et~al.} 2017, \mnras, 466, 4193,
  \dodoi{10.1093/mnras/stw3301}

\bibitem[{{Ueda} {et~al.}(2014){Ueda}, {Akiyama}, {Hasinger}, {Miyaji}, \&
  {Watson}}]{Ueda2014ApJ...786..104U}
{Ueda}, Y., {Akiyama}, M., {Hasinger}, G., {Miyaji}, T., \& {Watson}, M.~G.
  2014, \apj, 786, 104, \dodoi{10.1088/0004-637X/786/2/104}

\bibitem[{Virtanen {et~al.}(2020)Virtanen, Gommers, Oliphant, Haberland, Reddy,
  Cournapeau, Burovski, Peterson, Weckesser, Bright, {van der Walt}, Brett,
  Wilson, Millman, Mayorov, Nelson, Jones, Kern, Larson, Carey, Polat, Feng,
  Moore, {VanderPlas}, Laxalde, Perktold, Cimrman, Henriksen, Quintero, Harris,
  Archibald, Ribeiro, Pedregosa, {van Mulbregt}, \& {SciPy 1.0
  Contributors}}]{2020SciPy-NMeth}
Virtanen, P., Gommers, R., Oliphant, T.~E., {et~al.} 2020, Nature Methods, 17,
  261, \dodoi{10.1038/s41592-019-0686-2}

\bibitem[{{Wang} \& {Loeb}(2016)}]{Wang2016NatPh..12.1116W}
{Wang}, X., \& {Loeb}, A. 2016, Nature Physics, 12, 1116,
  \dodoi{10.1038/nphys3837}

\bibitem[{{Wojaczy{\'n}ski} \&
  {Nied{\'z}wiecki}(2017)}]{Wojaczy2017ApJ...849...97W}
{Wojaczy{\'n}ski}, R., \& {Nied{\'z}wiecki}, A. 2017, \apj, 849, 97,
  \dodoi{10.3847/1538-4357/aa8f9d}

\bibitem[{{Wojaczy{\'n}ski} {et~al.}(2015){Wojaczy{\'n}ski}, {Nied{\'z}wiecki},
  {Xie}, \& {Szanecki}}]{Wojaczy2015A&A...584A..20W}
{Wojaczy{\'n}ski}, R., {Nied{\'z}wiecki}, A., {Xie}, F.-G., \& {Szanecki}, M.
  2015, \aap, 584, A20, \dodoi{10.1051/0004-6361/201526621}

\bibitem[{Zdziarski {et~al.}(1996)Zdziarski, Johnson, \&
  Magdziarz}]{Zdziarski10.1093}
Zdziarski, A.~A., Johnson, W.~N., \& Magdziarz, P. 1996, Monthly Notices of the
  Royal Astronomical Society, 283, 193, \dodoi{10.1093/mnras/283.1.193}

\end{thebibliography}
\bibliographystyle{aasjournal}

\end{document}